\documentclass[nofootinbib,amsmath,amssymb,showpacs,showkeys,aps,reprint]{revtex4-1}
\usepackage[utf8,latin1]{inputenc}
\usepackage{graphicx}
\usepackage{dcolumn}
\usepackage[dvipsnames]{xcolor}
\usepackage[T1]{fontenc}

\usepackage{mathrsfs}  
\usepackage{cases}
\usepackage{comment} 
\usepackage{bm}
\usepackage{academicons}
\usepackage{mathtools, nccmath}
\usepackage{fancyhdr}
\usepackage{tikz,xcolor}
\newcommand{\qm}[1]{``#1''}
\usepackage{tensor}
\usepackage[normalem]{ulem}
\usepackage{lipsum}
\usepackage{soul}
\usepackage{cancel}
\usepackage{stackengine,scalerel}
\usepackage{mathabx}
\usepackage{hyperref}
\usepackage{tabularx}
\hypersetup{colorlinks, linkcolor={red},citecolor={blue},urlcolor={blue}}

\def\Rbol{{\stackrel{\circ}{R}}{}} 
\def\Gbol{{\stackrel{\circ}{G}}{}} 
\def\Gammabol{{\stackrel{\circ}{\Gamma}}{}}
\def\Gammafw{{\stackrel{FW}{\Gamma}}{}}
\def\nablabol{{\stackrel{\circ}{\nabla}}{}}
\def\nablafw{{\stackrel{FW}{\nabla}}{}}
\def\abol{{\stackrel{\circ}{a}}{}}
\def\Dbol{{\stackrel{\circ}{D}}{}}

\def\Pperp{\mathop{\mathcal{P}_{\perp }}\nolimits}
\def\Pparll{\mathop{\mathcal{P}_{\parallel }}\nolimits}
\def\PperpGR{\mathop{\mathcal{\stackrel{\circ}{P}}_{\perp }}\nolimits}
\def\PparllGR{\mathop{\mathcal{\stackrel{\circ}{P}}_{\parallel }}\nolimits}
\def\Fperp{\mathop{\mathcal{F}_{\perp}}\nolimits}
\def\Fparll{\mathop{\mathcal{F}_{\parallel}}\nolimits}
\definecolor{lime}{HTML}{A6CE39}
\DeclareRobustCommand{\orcidicon}{
	\begin{tikzpicture}
	\draw[lime, fill=lime] (0,0) 
	circle [radius=0.16] 
	node[white] {{\fontfamily{qag}\selectfont \tiny ID}};
	\draw[white, fill=white] (-0.0625,0.095) 
	circle [radius=0.007];
	\end{tikzpicture}
	\hspace{-2mm}
}

\def\nn{\nonumber} 
\newcommand{\dd}{{\rm d}}

\newcommand\F{\mathcal{F}}

\fancypagestyle{plain}{%
  \fancyhf{}
  \fancyfoot[C]{\iffloatpage{}{\thepage}}
  }
\foreach \x in {A, ..., Z}{%
	\expandafter\xdef\csname orcid\x\endcsname{\noexpand\href{https://orcid.org/\csname orcidauthor\x\endcsname}{\noexpand\orcidicon}}
}
\begin{document}

\title[Equivalence Principle violation in   metric-affine  gravity and finite-temperature effects]{Equivalence Principle violation in   metric-affine  gravity and finite-temperature effects}

\author{Emmanuele Battista\orcidA{}$^{1}$\vspace{0.5cm}}\email{ebattista@lnf.infn.it} 
\author{Roberto Campagnola\orcidD{}$^{1}$\vspace{0.5cm}} \email{roberto.campagnola@lnf.infn.it}
\author{Salvatore Capozziello\orcidB{}$^{2,3,4}$\vspace{0.5cm}} \email{capozziello@na.infn.it}
\author{Giuseppe Fiorillo\orcidC{}$^{5}$\vspace{0.5cm}} \email{giufiorillo@unisa.it}

\affiliation{$^1$ Istituto Nazionale di Fisica Nucleare, Laboratori Nazionali di Frascati, 00044 Frascati, Italy\\
$^2$ Dipartimento di Fisica ``Ettore Pancini'', Complesso Universitario 
di Monte S. Angelo, Universit\`a degli Studi di Napoli ``Federico II'', Via Cinthia Edificio 6, 80126 Napoli, Italy,\\
$^3$ Istituto Nazionale di Fisica Nucleare, Sezione di Napoli, Complesso Universitario 
di Monte S. Angelo, Via Cinthia Edificio 6, 80126 Napoli, Italy,\\
$^4$  Scuola Superiore Meridionale, Largo San Marcellino 10, 80138 Napoli, Italy,\\
$^5$ Dipartimento di Fisica ``E.R. Caianiello'', Universit\`a degli Studi di Salerno, Via Giovanni Paolo II 132, 84084 Fisciano (SA), Italy}

\date{\today}

\begin{abstract}

Possible violations of the equivalence principle are investigated  within the framework of metric-affine  gravity and   their connection to finite-temperature effects are highlighted. Thermal  corrections to particle dynamics, originally derived in a quantum-field-theory  setting, can be evaluated in a purely Riemannian framework  and lead to a  shift in the gravitational-to-inertial mass ratio. We show that the ensuing departure from universality of free fall can be also formulated in metric-affine gravity, where the presence of the non-metricity tensor  modifies the Newtonian law in a way that closely parallels the finite-temperature scenario.
Furthermore, we introduce a generalized Fermi-Walker derivative adapted to non-Riemannian contexts, which naturally reveals that no orthonormal tetrad can be propagated along an observer worldline. Although metric-affine gravity admits a pointwise realization of the Einstein equivalence principle in its gauge-theoretic, elementary-matter form, the new operator offers a direct geometric signature that this principle, in its modern formulation, is not retained in general.  Potential tests of the analyzed effects  are also discussed.

\end{abstract}

\maketitle
%\tableofcontents

\section{Introduction}

The cornerstones of Einstein theory are the principles of relativity  and general covariance, together with the equivalence principle (EP) \cite{romano2019_book_GR}. The first asserts that \emph{all} observers have equal status in describing physical phenomena, a condition   closely connected with the  second principle, which states that the fundamental laws of physics can be formulated in tensor form on a spacetime manifold $\mathcal{M}$; the  EP,  in its Einstein formulation,  affirms that gravitational effects  can be locally eliminated in a freely falling frame. Together, these tenets led Einstein to formulate general relativity (GR) as a  geometric theory  on a four-dimensional (pseudo-)Riemannian manifold $\mathcal{M}$, endowed with a metric  tensor  $\tensor{g}{_{\mu\nu}}$ and  the symmetric, metric-compatible Levi-Civita connection.

Historically, the EP  appeared in  progressively refined versions, reflecting  the deepening of our understanding of gravitation \cite{Lebed:2020ros}. The Newtonian EP already recognized the universality of free fall through the equivalence of inertial and gravitational mass. The weak EP sharpens this by stating that all test bodies, regardless of their composition or internal structure, fall with the same acceleration in a given gravitational field. The \emph{modern formulation} of the Einstein EP further requires the local Lorentz invariance and the local position invariance, ensuring that non-gravitational experiments yield the same outcomes in any freely falling frame. Finally, the strong EP extends these statements to include also self-gravitating bodies.

By adopting the post-Newtonian approximation,  GR compliance  with the  strong EP has been explicitly confirmed up to the second post-Newtonian order \cite{Mitchell2007} (i.e., $O(c^{-4})$ beyond the Newtonian theory). However, it has been recently claimed that possible departures could emerge starting from the third post-Newtonian level, where inner-structure-dependent effects might potentially appear in the equations of motion (and the gravitational waveforms) of non-spinning compact bodies \cite{Will2025}. 

Empirical probes of the EP span laboratory and space experiments \cite{Will:2014kxa,Berti2015,GBAR2015,Viswanathan2017,Will-book2018,Belenchia2021}. Violations of the weak EP are parameterized by the E\"{o}tv\"{o}s ratio $\eta $, which quantifies fractional acceleration differences between test bodies of different composition. This parameter is stringently constrained by modern torsion-balance experiments \cite{Wagner2012} (e.g., E\"{o}tv\"{o}s, Dicke, Braginsky, and the E\"{o}t-Wash group), as well as by free-fall  and matter-wave interferometry \cite{Overstreet2017,Asenbaum2020,Tino2020} and by  space-based missions  such as MICROSCOPE \cite{Touboul2017,MICROSCOPE2022}. Additional bounds are expected from forthcoming projects, including  STEP \cite{Pereira2016}, Galileo Galilei \cite{Nobili2012}, STE-QUEST \cite{STE-QUEST2022}, and ACES \cite{Cacciapuoti2024}. Moreover, tests of the weak and strong EP are  carried out through lunar laser-ranging techniques \cite{Williams2005,Williams2012,Congedo2016}. Although these efforts    underpin metric theories of gravity in general, they  still leave room for potential new physics beyond GR.

Deviations from the EP  are predicted in several theoretical frameworks,   motivated by cosmology, quantum gravity, and unification schemes   \cite{Will:2014kxa,Berti2015,Tino2020}.  Non-metric theories introduce additional fields that couple differently to matter, leading to composition-dependent accelerations and violations of the weak EP \cite{March2023,Kraiselburd2019}. Scalar-tensor theories, such as Brans-Dicke and its generalizations \cite{Damour2010,Armendariz-Picon2011}, predict variations in the effective gravitational constant and strong-field phenomena like the Nordtvedt effect or spontaneous scalarization in neutron stars, signaling a breakdown of the strong EP \cite{Zhang2019,Damour1993,Kuntz2024}.  Vector-tensor and tensor-vector-scalar paradigms introduce preferred-frame or preferred-location effects, contravening local Lorentz invariance (i.e., the condition that experimental outcomes are independent of the velocity of  particle's freely falling  frame) and local position invariance (i.e., the requirement that empirical results are unaffected by the test body location) \cite{Yagi2013,Skordis2009,Famaey2011}. Models with changing fundamental constants \cite{Uzan2002,Uzan2010,Dzuba2024}, including varying-speed-of-light scenarios \cite{Bileska2024}, also challenge local position invariance by allowing spacetime-dependent physical constants. Furthermore,  the EP can be spoiled in  massive gravity and higher-order curvature theories  \cite{Belikov2012,Hiramatsu2012,Olmo2006}, as well as in string-inspired effective frameworks \cite{Mavromatos2022}.

EP breakdowns are also expected  at quantum level \cite{Lammerzahl1996}. More  generally, the Einstein EP is inconsistent with scenarios in which either matter or gravity exhibit a quantum behavior, and some proposals have been advanced  to extend this tenet to the quantum domain (see e.g. Refs. \cite{Lammerzahl1996,Zych2015,Seveso2017,Lebed:2020ros,Cepollaro2024} and references therein). Moreover, violations of the weak EP  occur in effective-field-theory models of  gravity, where particles with different spins experience distinct bending angles when propagating in a quantum-corrected background  geometry  \cite{Bjerrum-Bohr2014,Bjerrum-Bohr2016,Bai2016,Chi2019}.

A context that illustrates how quantum phenomena can lead to departures from the EP is represented by finite-temperature quantum field theory (QFT), which generalizes standard QFT to systems in thermal equilibrium at nonzero temperature \cite{Das1997,Kapusta:2007ftft-book,Khanna2009,Mustafa2022}. This framework  permits the simultaneous  investigation of quantum and thermal fluctuations, and  naturally predicts that free fall can become non-universal  due to a temperature-induced shift between the inertial and gravitational masses of a particle \cite{Donoghue:1984zs, Donoghue:1984PE}. It has been proved that this result can be recovered via  an approach based on a Riemannian geometry where  local Lorentz symmetry is broken by  the presence of a heat bath. This analysis was first carried out in a GR setting \cite{Gasperini:1987vb} and later extended  to curvature-based modified metric gravity models \cite{Blasone:2021phx}.

In this paper, we show that potential  EP violations related to temperature effects can be naturally explained within the wider context of metric-affine geometry (MAG). MAG provides a  gauge-theoretic formulation of the gravitational interaction analogous in spirit to  Yang-Mills theories, with the  non-Abelian unitary group of internal symmetries replaced by a spacetime symmetry group, i.e., the general affine group.  In this framework,   the metric-compatibility constraint is relaxed, and the affine connection and the metric are treated as independent variables. Consequently, the  connection  acquires extra components beyond the Levi-Civita part that lead to a richer geometric domain where geodesics and autoparallels no longer coincide. As we will see, these features make MAG a natural arena for exploring  EP departures in the dynamics of microstructured test bodies. Moreover, we complete our analysis by introducing a generalized Fermi-Walker (FW) transport rule   adapted to MAG, which provides a purely geometric demonstration of the Einstein EP breakdown in full generality. 

Specifically,   MAG retains the (field-theoretic) pointwise, elementary-matter version of the Einstein EP \cite{Von-Der-Heyde-1975,Hehl1976}, whereby, at a given spacetime point, the nongravitational laws for minimally coupled matter can be reduced to their special-relativistic form in a suitable local frame. By contrast,  what is not in general valid is the modern formulation  of the Einstein EP, which relies on the weak EP and the existence of local Lorentz frames extended along worldlines, and hence  involves  additional assumptions that are not purely pointwise but probe the behavior of matter and frames beyond a single spacetime event. This situation stems from the fact that non-metricity and its (nontrivial hypermomentum) couplings to matter obstruct the realization of local  Lorentz frames beyond a single spacetime point, so that orthonormal tetrads cannot be preserved along worldlines  and the usual local inertial-frame construction is  spoiled. Interestingly, this picture aligns  with recent works suggesting that possible departures from local Lorentz invariance are naturally captured in MAG by a nonvanishing non-metricity, as opposed to the non-geometric formulations based on  non-dynamical Lorentz-violating tensor fields commonly employed in the literature \cite{Obukhov2023-proc,Obukhov2024a}.

The paper is organized as follows. After reviewing  the key concepts of MAG, including its gauge-theoretic foundations, in Sec. \ref{sec:MAG},  in Sec. \ref{sec:finite-temperature} we set out gravity at finite temperature from both quantum and GR viewpoints and describe the ensuing  EP departures.  In Sec. \ref{Sec:EP-violation-in-MAG-static-background}, we demonstrate  that these EP violations inherently occur in MAG and outline their potential experimental probes. Then,   in Sec. \ref{sec:parallel transport and GFW}, we introduce our novel extended FW derivative, and   lastly, in Sec. \ref{Sec:Conclusion}, we draw our conclusions. 

\emph{Notations and conventions.} We use units where $\hbar$, $G$, $c$, and the Boltzmann constant $k_{\rm B}$ are set equal to one. The metric signature is $(-+++)$.   Round (respectively, square) brackets around a pair of indices stand for the usual symmetrization (respectively, antisymmetrization)
procedure, i.e., $A_{(\mu \nu)}=\tfrac{1}{2}\left(A_{\mu \nu}+A_{\nu \mu}\right)$ (respectively, $A_{[\mu \nu]}=\tfrac{1}{2}\left(A_{\mu \nu}-A_{\nu \mu}\right)$).

\section{Metric-affine geometry}\label{sec:MAG}

This section has a twofold purpose. On the one hand, it provides an overview of the fundamental aspects of MAG, and, on the other, it allows us to set our conventions. 

MAG involves  a transition from the standard Riemannian structure of GR to a non-Riemannian  scenario,   where  additional geometric degrees of freedom arise  by allowing the connection to possess both torsion and non-metricity. Such features are considered in Sec. \ref{Sec:torsion-non-metricity-curvature}, while  the kinematical implications of this  richer geometry are examined in Sec. \ref{Sec:MAG-kinematics}. Since metric-affine gravity represents a gauge field theory \emph{par excellence}, we conclude the section by outlining this framework in Sec. \ref{Sec:MAG-gauge-theory}.

\subsection{Torsion and  non-metricity }\label{Sec:torsion-non-metricity-curvature}

Riemannian geometry represents the mathematical arena of GR. In this context, the affine structure is  defined by the Levi-Civita connection 
\begin{align} \label{Levi-Civita-connection}
\tensor{\Gammabol}{^{\alpha}_{\mu\nu}}= \frac{1}{2}\tensor{g}{^{\alpha\lambda}}\left( \tensor{\partial}{_\mu}\tensor{g}{_{\lambda\nu}} + \tensor{\partial}{_\nu}\tensor{g}{_{\mu\lambda}}-\tensor{\partial}{_\lambda}\tensor{g}{_{\mu\nu}} \right)   \,, 
\end{align}
which is completely determined by the metric tensor $\tensor{g}{^{}_{\mu\nu}}$. Henceforth, Greek indices $\mu,\nu,\dots=0,1,2,3$ stand for holonomic spacetime coordinate indices, and  quantities framed in Riemannian geometry  are denoted with an overcircle. 

In the broader MAG context, the connection $\tensor{\Gamma}{^{\alpha}_{\mu\nu}}$ attains the general form \cite{Schouten1954}
\begin{equation} \label{connection with KandL}
    \tensor{\Gamma}{^{\alpha}_{\mu\nu}} = \tensor{\Gammabol}{^{\alpha}_{\mu\nu}} + \tensor{M}{^{\alpha}_{\mu\nu}} \,, 
\end{equation}
where the non-Riemannian piece $\tensor{M}{^{\alpha}_{\mu\nu}}$,  referred to as  distortion tensor,  satisfies
\begin{align}
 M_{\alpha[\mu \nu]}&=-\frac{1}{2}T_{\alpha \mu \nu}\,,   
 \nonumber \\
  M_{(\alpha \mu) \nu}&=-\frac{1}{2}Q_{ \nu \alpha \mu}\,,
\end{align} 
with 
\begin{align} 
\tensor{T}{^{\alpha}_{\mu\nu}} :=& 2\tensor{\Gamma}{^{\alpha}_{[\mu\nu]}}  \,, \label{torsion tensor} \\
 \tensor{Q}{^{}_{\lambda\mu\nu}} :=& \tensor{\nabla}{^{}_{\lambda}}\tensor{g}{^{}_{\mu\nu}} \,,  \label{non-metricity tensor} 
\end{align}
 the torsion    and the non-metricity tensors, respectively.  Geometrically, $\tensor{T}{^{\alpha}_{\mu\nu}}$  measures the failure of the closure of the parallelogram
made up of small displacement vectors and their parallel transports \cite{Nakahara2003}, while $ \tensor{Q}{^{}_{\lambda\mu\nu}}$ quantifies the lack of covariant conservation of the metric  \cite{Bahamonde_2023}.  

The definition of $\tensor{M}{^{\alpha}_{\mu\nu}}$ in Eq. \eqref{connection with KandL}  is particularly useful for two reasons. First, it allows to separate the  covariant derivative into   Riemannian and non-Riemannian parts:
\begin{align} \label{general covariant derivative}
    \tensor{\nabla}{^{}_{\lambda}}\tensor{X}{^{\alpha}_{}} &= \tensor{\partial}{^{}_{\lambda}}\tensor{X}{^{\alpha}_{}} + \tensor{\Gamma}{^{\alpha}_{\rho\lambda}}\tensor{X}{^{\rho}_{}}  \nonumber\\
    &= \tensor{\partial}{^{}_{\lambda}}\tensor{X}{^{\alpha}_{}} + \tensor{\Gammabol}{^{\alpha}_{\rho\lambda}}\tensor{X}{^{\rho}_{}} + \tensor{M}{^{\alpha}_{\rho\lambda}}\tensor{X}{^{\rho}_{}}  \nonumber\\
    &=: \tensor{\nablabol}{^{}_{\lambda}}\tensor{X}{^{\alpha}_{}} + \tensor{M}{^{\alpha}_{\rho\lambda}}\tensor{X}{^{\rho}_{}} \,;
\end{align}
second,   the Riemann curvature tensor 
\begin{equation} \label{curvature tensor}
    \tensor{R}{^{\alpha}_{\beta\mu\nu}} := \partial_\mu\tensor{\Gamma}{^{\alpha}_{\beta\nu}} 
    - \partial_\nu\tensor{\Gamma}{^{\alpha}_{\beta\mu}} 
    + \tensor{\Gamma}{^{\alpha}_{\mu\lambda}}\tensor{\Gamma}{^{\lambda}_{\nu\beta}}
    - \tensor{\Gamma}{^{\alpha}_{\nu\lambda}}\tensor{\Gamma}{^{\lambda}_{\mu\beta}} \,,
\end{equation}
 can be decomposed in the following way \cite{Bahamonde_2023}: 
\begin{align} \label{curvature decomposition}
    \tensor{R}{^{\alpha}_{\beta\mu\nu}} = &
    \tensor{\Rbol}{^{\alpha}_{\beta\mu\nu}} 
    + \tensor{\nablabol}{_{\mu}}\tensor{M}{^{\alpha}_{\beta\nu}}
    - \tensor{\nablabol}{_{\nu}}\tensor{M}{^{\alpha}_{\beta\mu}}   + \tensor{M}{^{\alpha}_{\lambda\mu}}\tensor{M}{^{\lambda}_{\beta\nu}} \nonumber \\
    &- \tensor{M}{^{\alpha}_{\lambda\nu}}\tensor{M}{^{\lambda}_{\beta\mu}},
\end{align}
the only symmetry being, in general,   the antisymmetry in the last two indices \cite{Capozziello-Ferrara2022,Annala2022,Bahamonde_2023,Capozziello-Ferrra2025,Battista2026g}.

\subsection{Kinematics }\label{Sec:MAG-kinematics}

Owing to its rich structure, important implications arise at the kinematical level in MAG.  Unlike in Riemannian geometries,  geodesics and autoparallels no longer coincide (see Sec. \ref{sec:geodesic vs autoprallel}), and the latter generally exhibit an acceleration term, as we will discuss in Sec. \ref{sec:anomalous-accel}. 

For further details regarding the kinematical aspects of MAG we refer the reader to Refs. \cite{Iosifidis:2018diy,Agashe:2023vsz,Agashe2023b}.

\subsubsection{Geodesic and autoparallel curves} \label{sec:geodesic vs autoprallel}

In GR, there is no distinction between geodesic and autoparallel structures, although they  have distinct conceptual and technical origins \cite{Hehl1976}. The geodesic is an extremal curve and its formulation thus  arises from a variational principle. On the other hand, the autoparallel is defined as the path over which a vector is   transported parallel to itself according to the connection $\tensor{\Gamma}{^{\alpha}_{\mu\nu}}$ of the manifold $\mathcal{M}$, thereby generalizing the straight line notion of  flat spacetime.

Based on its defining property, the geodesic equation takes the form
\begin{equation} \label{geodesics equation}
    \tensor{\Ddot{x}}{^\lambda} + \tensor{\Gammabol}{^{\lambda}_{\mu\nu}}\tensor{\Dot{x}}{^\mu}\tensor{\Dot{x}}{^\nu} = 0 \,,
\end{equation}
where the dot denotes from now on differentiation with respect to the affine parameter:
\begin{equation} \label{4velocity in mag}
    \tensor{\Dot{x}}{^{\mu}_{}} \equiv \frac{d\tensor{x}{^{\mu}_{}}}{ds}, 
\end{equation}
and it is clear that only the Levi-Civita connection \eqref{Levi-Civita-connection} is involved. In contrast, autoparallel equation 
\begin{equation} \label{autoparallel equation}
    \tensor{\Ddot{x}}{^\lambda} + \tensor{\Gamma}{^{\lambda}_{\mu\nu}}\tensor{\Dot{x}}{^\mu}\tensor{\Dot{x}}{^\nu} = 0 \,,
\end{equation}
depends on the full affine connection $\tensor{\Gamma}{^{\lambda}_{\mu\nu}}$. It follows from identity  \eqref{connection with KandL} that  Eq. \eqref{autoparallel equation} can be expressed as 
\begin{equation} \label{autoparallel equation v2}
    \tensor{\Ddot{x}}{^\lambda} + \tensor{\Gammabol}{^{\lambda}_{\mu\nu}}\tensor{\Dot{x}}{^\mu}\tensor{\Dot{x}}{^\nu} = -\tensor{M}{^{\lambda}_{\mu\nu}}\tensor{\Dot{x}}{^\mu}\tensor{\Dot{x}}{^\nu} \,,
\end{equation}
which could be seen as an initial indication of a potential EP violation within the MAG framework. We will come back to this point in Sec. \ref{Sec:pure-dilation}.

\subsubsection{The anomalous acceleration }\label{sec:anomalous-accel}

The existence  of a nonzero non-metricity tensor \eqref{non-metricity tensor} entails significant consequences for the spacetime structure, mainly because   care must be taken when evaluating  the covariant derivative of covariant or contravariant vectors (cf. Eq. \eqref{general covariant derivative}). 
In general, the main effects introduced by $\tensor{Q}{^{}_{\lambda\mu\nu}}$ are: (i) lengths and angles are typically not preserved during the parallel transport of vectors; (ii)  autoparallels exhibit an acceleration term, commonly referred to as  anomalous acceleration. 

To illustrate property (i),  consider the four-velocity vector \eqref{4velocity in mag}. Even along an autoparallel curve, where  $\tensor{\Dot{x}}{^{\lambda}}\tensor{\nabla}{^{}_{\lambda}} \tensor{\Dot{x}}{^{\mu}_{}} = 0$, its norm is not conserved like in GR, as one has 
\begin{align} \label{demonstration magnitude 4vel}
    \frac{D}{ds}(\tensor{g}{_{\mu\nu}}\tensor{\Dot{x}}{^{\mu}}\tensor{\Dot{x}}{^{\nu}}) &=  2\tensor{\Dot{x}}{^{\lambda}}(\tensor{\nabla}{^{}_{\lambda}} \tensor{\Dot{x}}{^{\mu}_{}})\tensor{\Dot{x}}{_{\mu}} + \tensor{\Dot{x}}{^{\lambda}_{}}\tensor{Q}{^{}_{\lambda\mu\nu}}\tensor{\Dot{x}}{^{\mu}_{}}\tensor{\Dot{x}}{^{\nu}_{}}  \nonumber\\
    &= \tensor{\Dot{x}}{^{\lambda}_{}}\tensor{Q}{^{}_{\lambda\mu\nu}}\tensor{\Dot{x}}{^{\mu}_{}}\tensor{\Dot{x}}{^{\nu}_{}}  \,,
\end{align}
where $\frac{D}{d s}(\cdot) = \tensor{\Dot{x}}{^{\alpha}}\tensor{\nabla}{^{}_{\alpha}} \left(\cdot\right) $ represents the covariant directional derivative along $\tensor{\Dot{x}}{^{\alpha}}$. At this stage, it is thus convenient to  parametrize  the magnitude of $\tensor{\Dot{x}}{^{\mu}_{}}$ via  \cite{Agashe:2023vsz}
\begin{equation} \label{magnitude 4vel}
    \tensor{\Dot{x}}{^{\mu}_{}}\tensor{\Dot{x}}{^{}_{\mu}} = - \phi(\tensor{x}{^\rho}) \,,
\end{equation}
where  $\phi$ is a positive-definite function of the spacetime point $x^\rho$.

The point (ii) concerns some subtleties in the expression of the four-acceleration. In GR, the commutativity between the metric and the covariant derivative implies that there is no distinction as to whether the   four-acceleration definition involves  the contravariant or covariant form of the four-velocity. In  metric-affine gravity, however,  this  equivalence no longer holds,  and it is possible to define two independent types of acceleration \cite{Iosifidis:2018diy,Agashe:2023vsz,Capozziello2024a}: the  proper acceleration
\begin{equation} \label{proper acceleration in mag}
    \tensor{A}{^{\mu}_{}} := \tensor{\Dot{x}}{^{\lambda}_{}} \tensor{\nabla}{^{}_{\lambda}} \tensor{\Dot{x}}{^{\mu}_{}},
\end{equation}
and the anomalous acceleration
\begin{equation} \label{hyper acceleration in mag}
    \tensor{\Tilde{a}}{_{\mu}^{}} := \tensor{\Dot{x}}{^{\lambda}_{}} \tensor{\nabla}{^{}_{\lambda}} \tensor{\Dot{x}}{_{\mu}^{}} \,,
\end{equation}
which differ owing  to the non-metricity. Specifically,   
\begin{align} \label{hyper acceleration 2}
    \tensor{\Tilde{a}}{_{\mu}^{}} &= \tensor{\Dot{x}}{^{\lambda}_{}} \tensor{\nabla}{^{}_{\lambda}} (\tensor{g}{_{\mu\nu}}\tensor{\Dot{x}}{^{\nu}}) = \tensor{g}{_{\mu\nu}}\tensor{A}{^{\nu}} + \tensor{Q}{_{\lambda\mu\nu}}\tensor{\Dot{x}}{^{\lambda}}\tensor{\Dot{x}}{^{\nu}} \nonumber\\
    &= \tensor{A}{_{\mu}} + \tensor{Q}{_{\lambda\mu\nu}}\tensor{\Dot{x}}{^{\lambda}}\tensor{\Dot{x}}{^{\nu}} \,,
\end{align}
clearly showing that $ \tensor{\Tilde{a}}{_{\mu}^{}} \neq g_{\mu \nu}  \tensor{A}{^{\nu}_{}} $. Therefore, we see that in MAG, 
autoparallels, i.e.,  curves for which $A^\mu=0$,  have, in general, a nonzero  acceleration that is represented by  the anomalous term \eqref{hyper acceleration 2}. 

An additional key difference with respect to Riemannian geometry is that both proper and anomalous  accelerations are  not orthogonal to the  four-velocity, as it follows from the above formulas that 
\begin{subequations}
\label{accel-veloc-product-MAG}
\begin{align}
\tensor{A}{^\mu}\tensor{\Dot{x}}{_{\mu}} &= \frac{1}{2}\left( -\Dot{\phi} - \tensor{Q}{_{\lambda\mu\nu}}\tensor{\Dot{x}}{^{\lambda}}\tensor{\Dot{x}}{^{\mu}}\tensor{\Dot{x}}{^{\nu}} \right) \,,  \label{MAG-relation-1}\\
\tensor{\Tilde{a}}{^\mu}\tensor{\Dot{x}}{_{\mu}} &= \frac{1}{2}\left( -\Dot{\phi} + \tensor{Q}{_{\lambda\mu\nu}}\tensor{\Dot{x}}{^{\lambda}}\tensor{\Dot{x}}{^{\mu}}\tensor{\Dot{x}}{^{\nu}} \right) \,,
\end{align}
\end{subequations}
which in turn imply
\begin{subequations} \label{+- 4-acc}
\begin{align}
(\tensor{A}{^{\mu}_{}} + \tensor{\Tilde{a}}{^{\mu}})\tensor{\Dot{x}}{_{\mu}} &= -\Dot{\phi} \,, \label{scalar-proper-hyper-velocity-1} \\
(\tensor{A}{^{\mu}_{}} - \tensor{\Tilde{a}}{^{\mu}})\tensor{\Dot{x}}{_{\mu}} &= - \tensor{Q}{_{\lambda\mu\nu}}\tensor{\Dot{x}}{^{\lambda}}\tensor{\Dot{x}}{^{\mu}}\tensor{\Dot{x}}{^{\nu}} \,,
\end{align}
\end{subequations}
where we have exploited Eq. \eqref{magnitude 4vel}. Bearing in mind Eq. \eqref{MAG-relation-1}, one finds that the magnitude $\phi$  admits the following integral expression along an autoparallel curve $C_A$:
\begin{equation} \label{magnitude as metricitiy integral}
    \phi(\tensor{x}{^{\rho}}) = -\int_{C_A}  \tensor{Q}{_{\lambda\mu\nu}}\tensor{\Dot{x}}{^{\lambda}}\tensor{\Dot{x}}{^{\mu}}\tensor{\Dot{x}}{^{\nu}} \,ds \;+ k, \qquad (k \in \mathbb{R}) \,,
\end{equation}
which clearly shows that only in the Riemannian case,  where $\tensor{Q}{_{\lambda\mu\nu}}=0$, the  norm of the four-velocity  remains constant and can be consistently renormalized.

The concepts discussed in this section will be particularly useful for our analysis of Sec. \ref{sec:parallel transport and GFW}, where we will set forth the generalized FW derivative and discuss the Einstein EP violation in MAG.

\subsection{Metric-affine gauge theory of gravity}\label{Sec:MAG-gauge-theory}

Metric-affine gravity arises as a gauge-theoretic formulation of gravitation, where spacetime geometry is described in terms of gauge potentials associated with local spacetime symmetries.

The gauge principle plays a central role in the modern description of fundamental interactions. Historically \cite{ORaifeartaigh1997-book,Cao2019}, inspired by the pioneering  works of Weyl and later by the development of Yang-Mills theories for  (internal) unitary symmetry groups, the  gauging formalism was extended to (external) spacetime symmetries by Utiyama. The  subsequent studies of Kibble and Sciama culminated in  the formulation of Poincar\'e gauge gravity, where the spacetime connection is endowed with a nontrivial torsion but vanishing non-metricity. The natural further step then consisted in extending the gauge pattern to accommodate a completely general affine connection (cf. Eqs. \eqref{connection with KandL}--\eqref{non-metricity tensor}). This led to  metric-affine gauge theory of gravity, which is  obtained by promoting the full affine group $A(4,\mathbb{R})$ to a local gauge symmetry of spacetime \cite{Hehl1995-MAG,Gronwald-Hehl1996,Gronwald1997,JimenezCano2021}. Since  $A(4,\mathbb{R})$ comprises both four-parameter translations  and general linear transformations $GL(4,\mathbb{R})$, such extension of the gauge group beyond the Lorentz subgroup $SO(1,3)$  can be interpreted physically as a possible departure from the local Lorentz invariance, which manifests in the fact that light cones cease to have the status of  absolute spacetime structures \cite{Obukhov2023-proc,Obukhov2024a}. In accordance with the gauge procedure,  a nontrivial non-metricity emerges and the Riemann-Cartan geometry underlying  Poincar\'e gauge gravity is extended to a MAG, where both the metric and the affine connection  naturally become independent dynamical variables. 

The main facets of the model are given in Sec. \ref{Sec:gauge-affine-group}, while the role of the EP and its relation with the gauge principle are discussed in Secs. \ref{Sec:gauge-principle-EP} and \ref{Sec:gauge-principle}.

\subsubsection{The gauging procedure of the affine group}\label{Sec:gauge-affine-group}

The key idea characterizing metric-affine gravity follows from the gauge-theoretic principle  that  the description of physics should be independent of the choice of  reference frames used to express the dynamical variables of the theory. 

The starting point then consists in requiring that physical matter fields  be decomposed with respect to arbitrary affine frames, a construction that enables  suitable $GL(4,\mathbb{R})$ representations to act on their components. An affine frame  at an event  $x \in \mathcal{M}$  is made up of a linear frame $\boldsymbol{e}_A$ of the local tangent space $T_x\mathcal{M}$ (capital Latin indices $A,B\dots$ denote internal affine-frame components), together with  an associated point $p$ living in the affine tangent space $A_x\mathcal{M}$ \cite{Gronwald1997}.  In this framework, the \emph{gauge principle} thus encodes the independence of physical results under changes of affine frames $\left(\boldsymbol{e}_A,p\right)$, with affine transformations being identified as gauge transformations relating equivalent frames (in more geometric terms,  the affine  bundle, defined as the collection of all affine tangent spaces $A_x\mathcal{M}$, is the principal bundle of the gauge procedure and its  structure group is the affine group $A(4,\mathbb{R})$). Enforcing this equivalence  entails the introduction of a generalized affine connection $(\tensor{\boldsymbol{\Gamma}}{^{(L)}^A_B},\boldsymbol{\Gamma}^{(T)A})$, which maps infinitesimally neighboring affine tangent spaces. The $GL(4,\mathbb{R})$-valued linear part $\tensor{\boldsymbol{\Gamma}}{^{(L)}^A_B}$ governs rotations and deformations of the basis vectors $\boldsymbol{e}_A \in T_x\mathcal{M}$, while the translational $\mathbb{R}^4$-valued piece $\boldsymbol{\Gamma}^{(T)A}$ accounts for the shift of the reference point $p \in A_x\mathcal{M}$.

Unlike  (internal) Yang-Mills theories, where  unitary symmetry groups act in internal spaces, (external) gauge-gravity models require a \emph{soldering} mechanism that relates  the (internal) affine-geometric gauge structure to the (external) spacetime manifold \cite{Aldrovandi1996-book}. This is achieved by identifying 
the affine tangent space $A_x \mathcal{M}$ with the physical tangent space $T_x \mathcal{M}$, thereby permitting  the conversion of  affine-frame indices $A,B\dots$ into spacetime ones. As a result, the translational gauge invariance in the affine space is no longer realized as an
independent (internal) symmetry, but a local one-to-one correspondence is established with spacetime diffeomorphisms, which can then be interpreted as locally gauged translations. Another effect of soldering is that the translational component of the generalized affine connection   can be used to construct the translation-invariant anholonomic coframe $\boldsymbol{\vartheta}^a$, where $a,b,\dots=\hat{0},\hat{1},\hat{2},\hat{3}$ are hereafter anholonomic spacetime tangent-frame indices. This object  is related to the coordinate basis one-form $\boldsymbol{{\rm d}} x^\mu  $ via $\boldsymbol{\vartheta}^a = \tensor{e}{_\mu^a} \boldsymbol{\dd} x^\mu$ and is commonly called soldering one-form \cite{Aldrovandi1996-book}, the dual frame being denoted by $\boldsymbol{e}_a$. Moreover, the linear part of the affine connection becomes the $GL(4,\mathbb{R})$-valued connection $\boldsymbol{\Gamma}^a{}_{b}$ and  underlies the changes of frames under general linear transformations. Together, the pair $\left(\boldsymbol{\vartheta}^a,\boldsymbol{\Gamma}^a{}_{b}\right)$  constitutes  the true gauge potentials  of metric-affine gravity, endowed with a genuine dynamical role.

A separate procedure is necessary for  the metric, since it is not determined within the affine gauging scheme  but emerges naturally only after restricting the gauge symmetry to an affine-orthogonal subgroup \cite{Gronwald1997}\footnote{It has been proposed that the metric has a Goldstone nature in MAG; see Ref. \cite{Tresguerres2000} for further details.}. This means that in MAG the existence of a dynamical and independent local  metric structure  can  only be \emph{postulated}. Assume that  the  holonomic components  $g_{\mu \nu}$ with respect to some coordinate basis  have been specified;  these can then be expressed in terms of the coframe components $\tensor{e}{_\mu^a}$ via the following approach.  At a given point $x \in \mathcal{M}$ and in a suitable gauge  (referred to as orthonormal gauge \cite{Hehl1995-MAG}), which we denote by a star symbol, an orthonormal frame $\bar{\boldsymbol{e}}_a$ can be selected such that 
\begin{subequations}
\label{MAG-orthonormality}
\begin{align}
\left. \boldsymbol{g} \right \vert_x = g_{\mu \nu} \boldsymbol{\dd} x^\mu \otimes \boldsymbol{\dd} x^\nu \overset{\star}{=} \eta_{ab} \boldsymbol{\vartheta}^a \otimes \boldsymbol{\vartheta}^b, \label{MAG-orthonormality-1}
\end{align}
\text{which implies}
\begin{align}
\left. g_{\mu \nu} \right \vert_x \overset{\star}{=}  \tensor{\bar{e}}{_\mu^a} \tensor{\bar{e}}{_\nu^b} \eta_{ab}, \label{MAG-orthonormality-2}
\end{align}
\end{subequations}
with $\eta_{ab} ={\rm diag}\left(-1,1,1,1\right)$. Therefore,  the coframe $\boldsymbol{\vartheta}^a$  fixes the metric at a single spacetime point and in a specific gauge, where it can be chosen orthonormal \cite{Hehl1995-MAG,Gronwald1997}. 

The above result stems from simple linear-algebra arguments. Indeed, rather than Eq. \eqref{MAG-orthonormality-2}, the most general formula providing the transition between the anholonomic and holonomic formulation is 
\begin{align}
   g_{\mu \nu} =   \tensor{e}{_\mu^a} \tensor{e}{_\nu^b} g_{ab}, 
\end{align}
with $g_{ab}=g\left(\boldsymbol{e}_a,\boldsymbol{e}_b\right)$   representing, at an arbitrary point $x \in \mathcal{M}$,   a symmetric and non-degenerate matrix. It is  then  always possible to perform a suitable local $GL(4,\mathbb{R})$ deformation  to a frame where the metric is diagonal with eigenvalues  normalized to $\pm 1$. In this new orthonormal basis, one can thus write $g\left(\bar{\boldsymbol{e}}_a,\bar{\boldsymbol{e}}_b\right)=: \eta_{ab}$, and the matrices $\tensor{\bar{\Lambda}}{^a_b}(x) \in GL(4,\mathbb{R})$ that preserve orthonormality belong to the Lorentz subgroup $SO(1,3)$ of $GL(4,\mathbb{R})$. Hence, orthonormality is a pointwise gauge choice in MAG, and,  due to the presence of  non-metricity, it is not preserved under parallel transport. This situation differs from models in which the local gauge group is reduced to  $SO(1,3)$, where $\boldsymbol{\vartheta}^a$ becomes orthonormal in every allowed gauge at each spacetime point. As we will see in Sec. \ref{Sec:gauge-principle-EP}, relations \eqref{MAG-orthonormality} are crucial for the EP realization, as they represent the metric-affine analogue of the pointwise orthonormalization underlying the construction of local inertial frames.

Once the concept of gauge potentials is enlarged to encompass, apart from $\boldsymbol{\vartheta}^a$ and $\boldsymbol{\Gamma}^a{}_{b}$,  also the metric $g_{ab}$, the geometry of the spacetime is characterized accordingly by three independent field strengths: the torsion two-form  $\boldsymbol{T}^a = \boldsymbol{\dd} \boldsymbol{\vartheta}^a + \tensor{\boldsymbol{\Gamma}}{^a_b} \wedge \boldsymbol{\vartheta}^b$,  the curvature two-form $\boldsymbol{R}^a{}_b = \boldsymbol{\dd} \boldsymbol{\Gamma}^a{}_b - \boldsymbol{\Gamma}^a{}_c \wedge \boldsymbol{\Gamma}^c{}_b$,  and the non-metricity one-form $\boldsymbol{Q}_{ab} = \boldsymbol{\dd} g_{ab} - \left(\boldsymbol{\Gamma}^a{}_c g_{cb}+\boldsymbol{\Gamma}^b{}_c g_{ca}\right)$ (their holonomic counterparts are given in Eqs. \eqref{torsion tensor},  \eqref{curvature tensor}, and \eqref{non-metricity tensor}, respectively).  The field strengths enter the $A(4,\mathbb{R})$-gauge-invariant gravitational Lagrangian and are sourced by the canonical energy-momentum, hypermomentum, and symmetric (metric) energy-momentum tensors, which represent the matter currents of metric-affine gravity. These satisfy a set of generalized covariant Noether identities stemming from the  invariance of the matter Lagrangian under local diffeomorphisms (i.e.,  the translation symmetry after soldering) and  local $GL(4,\mathbb{R})$ transformations.

For further details, we refer the reader to Refs. \cite{Hehl1995-MAG,Gronwald-Hehl1996, Gronwald1997,Blagojevic2012,Ponomarev2017-book,JimenezCano2021}, while some aspects regarding the dynamics of  hypermomentum-charged test bodies will be addressed more closely  in  Sec. \ref{Sec:pure-dilation}.

\subsubsection{The  equivalence principle }\label{Sec:gauge-principle-EP}

In this section, we  explain how  the Einstein EP, in its gauge-theoretic formulation, is realized in MAG. 

The starting point is that, for every event $x \in \mathcal{M}$, there exists a suitable anholonomic system where
\begin{subequations}
\label{MAG-EP}
\begin{align}
\left. g_{ab} \right \vert_x  &\overset{\star}{=}   \eta_{ab},  \label{MAG-EP-1}
\\
\left. \boldsymbol{\Gamma}^a{}_{b} \right \vert_x   &\overset{\star}{=} 0. \label{MAG-EP-2}
\end{align}    
\end{subequations}
This is a purely pointwise gauge statement, following from the local $GL(4,\mathbb R)$  freedom of metric-affine gravity. Indeed, under a local linear transformation,  $\boldsymbol{e}'_a=\Lambda^b{}_a  \boldsymbol{e}_b$ and  $\boldsymbol{\vartheta}'{}^a=(\Lambda^{-1})^a{}_b \boldsymbol{\vartheta}^b$, so that the  transformed metric   at $x \in \mathcal{M}$ depends only on $\Lambda(x)$ \cite{Hehl1995-MAG}. On the other hand,  the connection one-form $\boldsymbol{\Gamma}^a{}_{b}$ behaves inhomogeneously, and its transformation law at $x$ also involves   $\partial_\mu \Lambda (x)$. This can be  seen by first writing   locally  $\boldsymbol{\Gamma}=\Gamma_\mu \boldsymbol{\dd} x^\mu$, where  each $\Gamma_\mu$ is a $4 \times 4$ matrix (i.e., $\left(\Gamma_\mu\right)^a{}_b$); then, under a local linear gauge  transformation, one finds 
$\Gamma'_\mu= \Lambda^{-1} \Gamma_\mu \Lambda + \Lambda^{-1} \partial_\mu \Lambda $ \cite{Hehl1995-MAG}. Thus, to obtain Eq. \eqref{MAG-EP},  one first chooses $\Lambda (x)$ in such a way that $g_{ab}(x)$  is brought to Minkowski form, as we saw in Eq. \eqref{MAG-orthonormality}, and then fixes $\partial_\mu \Lambda (x)$ so that the two terms in the transformed connection cancel at $x$, yielding $\boldsymbol{\Gamma}'{}^a{}_b(x)\overset{\star}{=}0$. 

The simultaneous trivialization \eqref{MAG-EP} of the metric  and  connection encodes the pointwise elementary-matter version of the Einstein EP, which asserts that locally the properties of special-relativistic matter in a noninertial frame  cannot be distinguished from the properties of the same matter in a corresponding gravitational field \cite{Von-Der-Heyde-1975,Hehl1976}. Consequently, the local action of a gravitational field on elementary matter can be simulated in flat Minkowski spacetime by going over to a noninertial frame of reference and, conversely, in curved spacetime there always exists the gauge freedom to locally eliminate the effects of a  true gravitational field. 

This gauge-theoretic approach to the EP reveals a close relationship with the minimal coupling  principle. Indeed,  the matter Lagrangian $\mathcal{L}_m$ minimally coupled to gravity involves $g_{ab}$ and $\boldsymbol{\vartheta}^a$, and  depends on the connection $\boldsymbol{\Gamma}^a{}_{b}$ only via the gauge covariant derivative of matter fields.  As a consequence, the EP entails that   $\mathcal{L}_m$,  along with  the ensuing local form of the nongravitational field equations,   matches at a given event its special-relativistic counterpart. However, the EP has a local (first-order) domain of applicability \cite{Hehl1976}, and hence higher-order (derivative-dependent) structures, such as conservation laws involving the (canonical) energy-momentum tensor, are not constrained to take their special-relativistic form, as they probe  nonlocal features of the fields. Hence, torsion and non-metricity can still affect matter through derivative terms and couplings to intrinsic degrees of freedom, such as spin and hypermomentum (as we will discuss in Sec. \ref{Sec:pure-dilation}).

Accordingly,  the EP does not imply that the field strengths  vanish, and, in particular,  $\boldsymbol{Q}_{ab}(x)  \overset{\star}{=}  \boldsymbol{\dd} g_{ab}(x)$. Therefore, one  cannot infer  from conditions \eqref{MAG-EP} the presence of a local (pseudo-)Euclidean structure associated with metric compatibility, consistently with the fact that the most general space with such a property is the Riemann-Cartan spacetime \cite{Von-Der-Heyde-1975,Hehl1976,Blagojevic2002-book}. This is the reason why the Einstein EP, in its  modern  sense relying on the realizability of local Lorentz frames  beyond a single point, is not valid in general in MAG, as we will prove in Sec. \ref{sec:parallel transport and GFW}.

\subsubsection{The role of the gauge principle} \label{Sec:gauge-principle}

From the  analysis of the previous section, it is clear that in MAG the pointwise gauge version of the Einstein EP is retained and that it is implemented mathematically via the minimal coupling  principle of matter to gravity \cite{Von-Der-Heyde-1975,Blagojevic2002-book}. This highlights  a structural relation with the gauge principle introduced in Sec.  \ref{Sec:gauge-affine-group}  \cite{Gronwald-Hehl1996,Gronwald1997,ORaifeartaigh1997-book,Cao2019}.

In fact, as pointed out before, the EP asserts the empirical fact that gravitational and inertial effects are locally indistinguishable. This  motivates the introduction of local inertial frames where gravitation can be  transformed away and, more generally, suggests a local description of spacetime geometry in terms of frames and connections.

From this standpoint, the gauge principle suggests a structurally analogous viewpoint, in which the choice of local  frames,  now extended to include spacetime frames, is likewise regarded as physically irrelevant. In this sense, spacetime frames can be treated on the same footing as internal frames in gauge theories. Metric-affine gravity implements this concept concretely  by claiming the local indistinguishability of local affine reference frames $\left(\boldsymbol{e}_A,p\right)$, in close analogy with Yang-Mills theory, where the corresponding indistinguishability concerns local bases in the representation space of the unitary group $SU(n)$. It then becomes natural to interpret changes of local frames as a local gauge symmetry.

Promoting such frame freedom to a local symmetry leads to the well-known gauge paradigm, namely  the requirement that invariance under a global symmetry persists under its local version. In metric-affine gravity, this tenet is applied to spacetime geometry itself and necessitates replacing ordinary derivatives by covariant ones via the  introduction of gauge connections, which describe how local frames are related at neighboring points and  precisely compensate for local frame changes \cite{Cao2019}.  This procedure corresponds to the minimal coupling prescription, and is consistent with  the Einstein EP  expectation that nongravitational  laws  reduce locally to their special-relativistic form via a suitable gauge choice of \emph{anholonomic} local frames, as encoded in identities \eqref{MAG-EP}.

\section{Gravitation at finite temperature} \label{sec:finite-temperature}

In its standard formulation, GR does not  account for temperature, although   the interplay between gravitation and thermal phenomena plays a crucial role in a variety of contexts \cite{Padmanabhan2009}. At finite temperature, matter no longer exists in vacuum but interacts with a heat bath, typically modeled as a background of photons or other relativistic particles. In this setup, EP violations are expected to arise, and   this issue can be tackled via  a formalism that permits  describing quantum fields in thermal equilibrium and their coupling to gravity. The first step toward this objective is provided by finite-temperature QFT, which we briefly outline  in Sec. \ref{Sec:Overview-FTQFT}. This  lays the basis for the analysis of gravity in thermal environments, which we explore in    Sec. \ref{subsec:gravitation at finite temperature}.

\subsection{Finite-temperature quantum field theory } \label{Sec:Overview-FTQFT}

Many physical phenomena of interest do not occur  in  vacuum, but in settings characterized by a nonzero temperature  $T$. For instance, in particle physics, this is the case in the description of the quark-gluon plasma state of matter \cite{Kapusta2003}. Moreover, temperature fulfills a non-negligible role for  phase transitions in models with spontaneously broken symmetries \cite{Dolan1973}, in the  early universe evolution  \cite{Rebhan:1994zw,Laine:2002zu,Saikawa2018}, in  astrophysical compact objects such as black holes \cite{Kovtun2004}, and many other fields (see Ref. \cite{Mustafa2022} for further details). 

The theoretical pattern to address these questions is represented by finite-temperature QFT, which extends conventional  QFT concepts \cite{Peskin1995,Weinberg1995} to systems in thermal equilibrium at $T \neq 0$ \cite{Peressutti1982,Khanna2009,Mustafa2022}. Three principal formalisms are employed to study thermal quantum fields: the imaginary-time \cite{Matsubara1955} and real-time approaches \cite{Schwinger1960b,Schwinger1960a,Keldysh1964, Lundberg:2020mwu} (devised by Matsubara,  and   Schwinger and Keldysh, respectively) and the thermo-field dynamics  method \cite{Umezawa1982} (introduced by Umezawa, Matsumoto, and Tachiki). For further details, we refer the reader to Refs. \cite{Das1997,Kapusta:2007ftft-book,Bellac2011,Laine2016}. 

Finite-temperature QFT  provides the necessary tools to compute thermal radiative corrections in particle propagators. In the next section, we discuss the gravitational implications of these effects.

\subsection{Equivalence Principle violation in finite-temperature gravity} \label{subsec:gravitation at finite temperature}

At finite temperature, particle masses acquire additional contributions  induced by interactions with the surrounding thermal bath. This phenomenon can be studied through  two complementary approaches: one based on a purely quantum framework  \cite{Donoghue:1984zs, Donoghue:1984PE} (see Sec. \ref{Sec:FTQFT-approach}), and another  rooted in a classical relativistic setting  \cite{Gasperini:1987vb} (see Sec. \ref{Sec:Gasperini-approach}).

\subsubsection{Quantum diagrammatic approach}\label{Sec:FTQFT-approach}

In their seminal papers \cite{Donoghue:1984zs, Donoghue:1984PE},  Donoghue and collaborators  exploited finite-temperature QFT tools to compute the corrections to the electron self-energy occurring in a low-temperature environment, where   
\begin{align}
  T \ll m_0,   \label{low-temperature-limit}
\end{align}
$m_0$ being the renormalized $T=0$  electron mass. In this regime, the background temperature  is mainly provided by a photon heat bath, with the effects of the electron-positron heat bath being exponentially suppressed by a factor proportional to $  e^{-m_0/T}$. Under these conditions, all calculations can be consistently carried out in the low-velocity-weak-field limit.

Due to the presence of  the nonzero  temperature \eqref{low-temperature-limit}, the particles' Feynman  propagator  receives  modifications involving the Bose-Einstein distribution function, which reflect the nontrivial structure of  thermal vacuum. As a consequence,  the inertial and gravitational masses, $m_i$ and $m_g$, of the electron renormalize   differently, as  one  finds, after an elaborate Feynman-diagram analysis \cite{Donoghue:1984zs,Donoghue:1984PE}, 
\begin{align} 
 m_i &= m_0 + \frac{\alpha \pi T^2}{3 m_0}\,, \label{inertial mass} \\
m_g &= m_0 - \frac{\alpha \pi T^2}{3 m_0} \,, \label{gravitational-massFTQFT}
\end{align}
with $\alpha$ the fine-structure constant. This implies that the electron acceleration in a gravitational field becomes mass-dependent,  thereby signaling the breaking of (at least) Newtonian EP  at $T \neq 0$. To leading order in the small parameter $T/m_0$,   the above formulas yield 
\begin{equation} \label{EP at finite T}
    \frac{m_g}{m_i} = 1 - \frac{2 \alpha \pi T^2}{3 m_0^2},
\end{equation}
a  result that can be attributed to the fact that  the heat-bath  singles out a preferred rest frame, leading to a lack of Lorentz invariance in the finite-temperature vacuum.

\subsubsection{Classical relativistic  approach}\label{Sec:Gasperini-approach}

A complementary analysis  was carried out by Gasperini \cite{Gasperini:1987vb}, who investigated thermal  effects on the gravitational acceleration of relativistic particles in the low-temperature limit \eqref{low-temperature-limit}, within  a weak and static gravitational field.  Unlike the previous investigation, which relies on diagrammatic techniques from finite-temperature QFT, this approach adopts a classical  gravity framework  where the  local Lorentz symmetry is broken in the  tangent space spanned by the tetrad field $\{ \tensor{e}{^\mu_a}\}$ (with  $a=\hat{0},\hat{1},\hat{2},\hat{3}$ the anholonomic frame index), while general covariance remains intact on the spacetime manifold $\mathcal{M}$. This construction gives rise to a quasi-Riemannian scenario, which however does not affect directly the underlying Riemannian geometry. Specifically, the Einstein equations retain their standard form 
\begin{equation}
{\Gbol}{^{\mu\nu}} = 8 \pi {\Theta}{^{\mu\nu}}\,, 
\end{equation}
with $\tensor{\Theta}{^{\mu\nu}}$ the generalized energy-momentum tensor  describing the effective source of gravity at finite temperature. 

In this pattern,  the ordinary  contracted Bianchi identity $    \tensor{\nablabol}{_\mu}\tensor{\Gbol}{^{\mu\nu}}=0$  can be still invoked, and  $\tensor{\Theta}{^{\mu\nu}}$ continues to be  covariantly conserved:
\begin{equation} \label{bianchi id in  gasperini}
\tensor{\nablabol}{_\mu}\tensor{\Theta}{^{\mu\nu}}=0 \, .
\end{equation}
At leading order in $T^2$, a detailed finite-temperature QFT calculation gives  the explicit form \cite{Donoghue:1984zs,Donoghue:1984PE}
\begin{equation}\label{Theta-tensor}
\tensor{\Theta}{^{\mu\nu}} = \tensor{T}{^{\mu\nu}} - \frac{2}{3}\alpha\pi\frac{T^2}{E^2}\tensor{e}{^\mu_{\hat{0}}}\tensor{e}{^\nu_{\hat{0}}}\tensor{T}{^{{\hat{0}{\hat{0}}}}} \,,
\end{equation} 
where $\tensor{T}{^{\mu\nu}}=\tensor{e}{^\mu_a} \tensor{e}{^\nu_b} T^{ab}$ is the standard particle stress-energy tensor minimally coupled to gravity, $e^\mu{}_{\hat{0}}$ picks out the heat-bath rest frame in the tetrad basis, and   the expression for the energy $E$ follows from the generalized  dispersion relation 
\begin{equation} \label{Emc2 at finite T}
    E = \sqrt{m_0^2 + |\vec{p}|^2 + \frac{2}{3}\alpha\pi T^2} \,, 
\end{equation} 
with $\vec{p}$ the   momentum three-vector of the source. The  tensor \eqref{Theta-tensor}  thus characterizes a matter distribution coupled to gravity in a generally covariant but not locally Lorentz-invariant manner: it transforms as a tensor under diffeomorphisms, but not as a scalar under local Lorentz transformations that rotates the tetrad frame in the local Minkowski tangent space \cite{Gasperini2021}.

Applying the Papapetrou multipole-expansion method  to the conservation law \eqref{bianchi id in  gasperini}, yields the following  modified worldline equation for a (spinless) test particle:
\begin{align} \label{autoparallel eq at finite T}
&\tensor{\Ddot{x}}{^\lambda} + \tensor{\Gammabol}{^{\lambda}_{\mu\nu}}\tensor{\Dot{x}}{^\mu}\tensor{\Dot{x}}{^\nu} 
\nonumber \\
&=  \frac{d}{ds}\left(\frac{2}{3}\alpha\pi\frac{T^2}{m E}\tensor{e}{^\lambda_{\hat{0}}}\right) + \frac{2}{3}\alpha\pi\frac{T^2}{m^2}\tensor{\Gammabol}{^\lambda_{\mu\nu}}\tensor{e}{^\mu_{\hat{0}}}\tensor{e}{^\nu_{\hat{0}}} \, ,
\end{align}
where the thermal contributions induce, at lowest order in $T^2/m^2$ ($m:= \sqrt{m_0^2 +2 \alpha \pi T^2/3}$),   a mass-dependent deviation from the  geodesic motion  that indicates  a breakdown of the universality of free fall.

A simple application  illustrating this situation is the case of a test body radially evolving in the Schwarzschild background sourced by a  mass $M$. In the weak-field approximation,  Eq. \eqref{autoparallel eq at finite T} leads to \cite{Gasperini:1987vb,Blasone:2021phx} 
\begin{equation} \label{radial motion at finite t}
    \Ddot{r} = -\frac{M}{r^2}\left(1 - \frac{2}{3}\alpha\pi\frac{T^2}{m^2}\right) \,,
\end{equation}
which shows that $m_g/ m_i$ matches the quantum-derived nonrelativistic formula \eqref{EP at finite T}. Remarkably, this result implies that, contrary to naive    thermodynamic expectations, which might suggest that thermal corrections to gravitational interactions vanish in the relativistic limit, the ratio $m_g/ m_i$ exhibits a constant thermal shift independent of the particle kinetic energy.

\section{Equivalence Principle violation from non-metricity}\label{Sec:EP-violation-in-MAG-static-background}

In the previous section, we  described how the inclusion of temperature  in gravitational  systems can spoil the EP. However, while in the formalism outlined in  Sec. \ref{subsec:gravitation at finite temperature} the underlying geometry is strictly Riemannian and departures from geodesic worldlines result from environment-induced breaking of local Lorentz symmetry in the tangent space, in this section,  we turn to the metric-affine framework, whose geometric richness goes beyond Riemannian structures and the obstruction to the operational realization of local Lorentz invariance can be naturally associated with the non-metricity. 

In Sec. \ref{Sec:pure-dilation}, we show that the presence of the  non-metricity tensor provides a natural arena for possible deviations from universality of free fall  by investigating the motion of a  test particle endowed with a dilation charge in a static and spherically symmetric background.   Then, in Sec. \ref{Sec:experiments},  we discuss how the resulting EP-violation signatures could be probed experimentally through future space missions and  high-precision laser ranging, and we highlight the  support offered by the astrometric data-analysis software called \emph{Planetary Ephemeris Program} ($PEP$).

\subsection{Pure-dilation motion in a static and spherically symmetric background}\label{Sec:pure-dilation}

The equations governing the trajectory of a microstructured and deformable test particle carrying spin, shear, and dilation charges in a generic non-Riemannian background\footnote{For recent investigations on the dynamical properties of generic MAGs, including black hole solutions, accretion processes,  stellar configurations and related astrophysical processes, as well as cosmological scenarios, see e.g. Refs. \cite{Bhatti2019,Maurya2022a,Maurya2023a,Mustafa2024a,Maurya2024a,Gohain2024,Yousaf2024,Nashed2025}. } follow from the Noether conservation laws  for the (canonical) energy-momentum  and  hypermomentum tensors,  defined, respectively, as
\begin{align}
\tensor{\Sigma}{^\mu_\nu} &:=\frac{\partial \mathcal{L}_m}{\partial\left(\nabla_\mu \psi\right)} \nabla_\nu \psi -\delta^\mu_\nu \mathcal{L}_m,
\\
\tensor{\Delta}{_\lambda^\mu^\nu} &:= -\frac{2}{\sqrt{-g}} \frac{\delta \left(\sqrt{-g}\mathcal{L}_m\right)}{\delta  \tensor{\Gamma}{^{\lambda}_{\mu\nu}}},\label{hyper-def}
\end{align}
$\psi$   denoting a generic matter field,   $\mathcal{L}_m$ the matter Lagrangian, and $g:= \det g_{\mu \nu}$.  Two  different but equivalent approaches can be pursued to obtain the dynamics of the test body: a covariant multipole-expansion method \cite{Puetzfeld2014}  or a  direct worldline derivation using Dirac-delta distributions (supplemented by the so-called convective \emph{ansatz})  \cite{Iosifidis-Hehl2023}. Moreover,  a Lagrangian formulation for a   hypermomentum-charged object  has been recently proposed in Ref. \cite{Iosifidis2025a} (see also Refs. \cite{Battista2022a,Battista2023a,DeFalco2023} for an analysis  involving the Weyssenhoff fluid in Einstein-Cartan theory, and  Ref. \cite{Battista2022b} for a study of  the semiclassical fermion motion in generic non-Riemannian backgrounds).  

As in any gauge theory, intrinsic properties of matter act as sources for the corresponding gauge fields. In MAG, these properties are encoded in the hypermomentum tensor \eqref{hyper-def}, which generalizes the notion of charge associated with internal symmetries. In particular, the spin, dilation, and shear currents represent additional fundamental characteristics of matter that couple to the affine connection, complementing the role of the energy-momentum tensor as the source of the metric field.

From a group-theoretic viewpoint, $\tensor{\Delta}{_\lambda^\mu^\nu}$ admits a \qm{fine structure} \cite{Obukhov2024a}, since it can be decomposed into three irreducible pieces under the general linear group $GL(4,\mathbb{R})$ \cite{Hehl1995-MAG}. The antisymmetric component corresponds to the intrinsic spin of matter and couples directly to spacetime torsion. The symmetric traceless portion represents the shear degrees of freedom, which capture distortions without changing the volume and interact with the traceless part of  non-metricity. Finally, the  trace term of  hypermomentum  describes the dilation charge (i.e., uniform changes of scale),  which couples to the Weyl sector of non-metricity: 
\begin{align}
    Q^\lambda:=g^{\mu \nu} \tensor{Q}{^\lambda_\mu_\nu}. 
\end{align}
This vector may be regarded as  providing the minimal geometric mechanism to violate local Lorentz invariance \cite{Obukhov2023-proc}. As discussed before, this phenomenon can be associated with the transition from the  group  $SO(1,3)$ to  $GL(4,\mathbb{R})$, which underlies the gauge paradigm of metric-affine gravity (see Sec. \ref{Sec:MAG-gauge-theory}). This observation is  relevant for the dynamics of test bodies, since the same non-metricity  components that encode this departure from local Lorentz symmetry also enter the equations of motion through the hypermomentum--non-metricity coupling \cite{Obukhov2023-proc,Obukhov2024a}.

If one focuses on a particle with inertial mass $m_i$ that is charged only under dilation, then the hypermomentum tensor takes the (convective) form \cite{Puetzfeld2014,Iosifidis-Hehl2023}
\begin{align}
    \tensor{\Delta}{_\lambda^\mu^\nu} = \delta^\mu_\lambda \Delta u^\nu, 
    \label{Weyl-vector-def}
\end{align}
and the worldline equation reads \cite{Puetzfeld2014,Iosifidis-Hehl2023}
\begin{align}
\frac{d^2 x^\nu}{d \tau^2} + \tensor{\Gammabol}{^{\nu}_{\mu\rho}} u^\mu u^\rho =-\frac{\Delta}{2m_i} u_\mu \partial^{[\mu} Q^{\nu]}, 
\label{dilation-motion}
\end{align}
where  we have employed the proper time $\tau$ to parametrize the dynamics and $u^\mu$ is the \emph{normalized} four-velocity, defined as
\begin{align}
u^\mu = \frac{d x^\mu}{d \tau} \equiv \dot{x}^\mu;   
\label{normalized-4-velocity-def}
\end{align}
moreover,  $\Delta u^\mu$ denotes the (conserved) dilation (vector) current, with the dilation charge density $\Delta$ being conserved along the particle path, i.e., $\dot{\Delta}=0$ \cite{Iosifidis-Hehl2023}. Notice that Eq. \eqref{dilation-motion} bears a formal resemblance to the general autoparallel equation of MAG (cf. Eq. \eqref{autoparallel equation v2}). 

Among the various hypermomentum charges appearing in MAG, the dilation coupling provides  the simplest and most transparent setting for isolating non-metricity-induced deviations from geodesic motion. We therefore focus on pure-dilation dynamics, and investigate  Eq. \eqref{dilation-motion} on the equatorial  $\theta=\pi/2$ plane of the static and spherically symmetric geometry
\begin{align}
d s^2 = -A(r) d t^2 +\frac{1}{B(r)} d r^2 + r^2 \left(d \theta^2 +  \sin^2 \theta \; d \varphi^2 \right); 
\label{static-spherically-symm-metric}
\end{align}
we also assume, consistently with the background symmetries, that the Weyl non-metricity vector \eqref{Weyl-vector-def} attains the form
\begin{align}
    Q^\mu(r)= \left(Q^t,Q^r,0,0\right). 
\end{align}

After a straightforward calculation,  we obtain for the temporal and radial components, respectively, 
\begin{subequations}
\begin{align}
&\Ddot{t} + \frac{A'}{A} \dot{t} \dot{r} = - \frac{d}{d \tau} \left(\frac{\Delta}{4m_i} Q^t\right),
\label{dilation-t-equation-1}
\\
& \Ddot{r} + \frac{B}{2} \left[\dot{t}^2 A' + \dot{r}^2 \left(B^{-1}\right)'\right]= - \frac{\Delta}{4m_i} AB \dot{t} \left(Q^t\right)',
\label{dilation-r-equation-1}
\end{align}
\end{subequations}
where the prime denotes the derivative with respect to the  radial variable $r$. Notice that the radial component $Q^r$ gives no contribution to the test particle trajectory in the static case. Bearing in mind  Eq. \eqref{dilation-t-equation-1}, Eq. \eqref{dilation-r-equation-1} can be expressed as  a total-$r$ derivative:
\begin{align}
    \frac{B}{2} \frac{d}{dr} \left(\frac{\dot{r}^2}{B}-A \dot{t}^2\right)=0,
\end{align}
whose solution is given by
\begin{align}
 \frac{\dot{r}^2}{B}-A \dot{t}^2=-1,
\end{align}
the constant factor on the right-hand side stemming from the normalization condition of the four-velocity \eqref{normalized-4-velocity-def}. With the help of the above relation, Eq. \eqref{dilation-r-equation-1} becomes the single radial equation
\begin{align}
\Ddot{r} &+\frac{B}{2} \Biggl[\frac{A'}{A}  +\frac{\dot{r}^2}{B}\left(\frac{A'}{A}-\frac{B'}{B}\right)+\frac{\Delta}{2m_i} \left(Q^t\right)' \sqrt{A+\frac{\dot{r}^2}{B}A}  \Biggr] 
\nonumber \\
&=0. 
\label{dilation-radial-final}
\end{align}

In order to  study Eq. \eqref{dilation-radial-final} in the weak-field-slow-motion regime, we suppose that the geometry \eqref{static-spherically-symm-metric}   refers to  a genuine gravitational field sourced by a mass $M$. Accordingly, we take  $g_{\mu \nu}$ to be asymptotically flat, so that for $r \to +\infty$ the functions $A(r)$ and $B(r)$ behave, to leading order, as
\begin{align}
A&=1+2 \Phi,
\nonumber \\
B^{-1}&= 1-2\Phi,
\end{align}
with $\Phi=-M/r$ the standard Newtonian potential (in geometrized units). Under these hypotheses, and  discarding higher-order   contributions,  Eq. \eqref{dilation-radial-final} reduces to 
\begin{align}
    \Ddot{r}=- \Phi' -\frac{\Delta}{4 m_i} \left(Q^t\right)',
\end{align}
showing that the additional (Lorentz-type) force  modifying the particle motion is entirely controlled by the derivative of the temporal component $Q^t$ of the Weyl vector. 

The above results highlight   the richness of  MAG framework, which allows us to explore several dynamical scenarios, including explicit tests of the EP. Indeed, let us  assume that in the weak-field limit $Q^t$ is proportional to 
$\Phi$:  
\begin{align}
Q^t = \chi \Phi,   
\end{align}
where the constant $\chi$ has the dimensions of an inverse length in $G=c=1$ units. 
In this way,   we arrive at 
\begin{align}  \label{dilation-MAG-weak-slow}
    \Ddot{r}=-\frac{M}{r^2} \left(1+\frac{\chi \Delta}{4 m_i}\right),
\end{align}
which  closely resembles   Eq. \eqref{radial motion at finite t}. We thus obtain the EP-violation relation
\begin{align}
    \frac{m_g}{m_i}=1+\frac{\chi \Delta}{4 m_i}, \label{mg-mi-ratio-MAG-1}
\end{align}
which yields
\begin{align}
    m_g = m_i + \frac{\chi \Delta}{4}. 
\end{align}
Equivalently, upon introducing  an explicit expansion around a reference (zero-order) inertial mass $m_0$ (e.g., the $T=0$ mass), we set $m_i=m_0+ O(\Delta)$ and then get, by neglecting terms $O\left(\Delta^2/m_0^2\right)$, 
 \begin{align}
     \frac{m_g}{m_i}=1+\frac{\chi \Delta}{4 m_0}. \label{mg-mi-ratio-MAG-3}
 \end{align}
 
Thus, the metric-affine pure-dilation force reproduces, at leading order, a Newtonian law with a small rescaling governed by the dilation charge density. This outcome is closely analogous to the finite-temperature results \eqref{inertial mass}--\eqref{EP at finite T}, which we recall have been derived in a Riemannian setting. Our analysis therefore demonstrates that EP violations stemming from the lack of Lorentz invariance of the thermal vacuum admit a natural reinterpretation using MAG formalism, where,  as we discussed before, departures from local Lorentz symmetry can be consistently  accounted for. Moreover, this  implies that non-Riemannian geometry naturally provides a  classical counterpart to the quantum effects responsible for the EP departure discussed in Sec. \ref{Sec:FTQFT-approach}.

Two crucial remarks are in order. First, the result \eqref{mg-mi-ratio-MAG-3} signals a violation of the Newtonian EP; however, if the effect described is not universal but composition or charge dependent, i.e., $\Delta/m_0$ differs between test bodies of different composition, then also the weak EP  is spoiled. 
Second, in the specific pure-dilation scenario  examined in this section, the Weyl-type non-metricity allows a redefinition of the affine parameter  into proper time, so that    a normalized four-velocity can be consistently adopted (cf. Eq. \eqref{normalized-4-velocity-def});  see Ref. \cite{Iosifidis-Hehl2023} for further details. However, in the broad  metric-affine arena,  non-metricity does not  preserve, in general,  the four-velocity norm, as we discussed in Sec. \ref{sec:MAG}.   For this reason, in  Sec. \ref{sec:parallel transport and GFW}, we will introduce an extended  FW derivative that permits  addressing the EP validity in  the  most general  MAG framework.

\subsection{Experimental probes via laser ranging and space missions}\label{Sec:experiments}

The above results could potentially be tested using laser-ranging methods and  space-based experiments.

Laser-ranging techniques currently provide one of the most sensitive empirical tests of the EP on lunar and planetary scales \cite{Williams2005,Congedo2016}. In particular, lunar laser ranging measures the round-trip time of laser pulses reflected by retro-reflectors on the Moon surface, yielding sub-centimeter (soon millimeter) constraints on the Earth-Moon dynamics  and on key fundamental observables \cite{DellAgnello2012a,Battista2015a,Battista2015b},  including  the weak and strong EP \cite{Williams2012}.  

The advent of next-generation retro-reflectors, most notably the \emph{MoonLIGHT} payload  integrated with a  double-axis pointing actuator for precise orientation, known as  MPAc (\emph{MoonLIGHT Pointing Actuator}),  has been designed to eliminate the pulse broadening caused by lunar librations that limit current Apollo and Lunokhod missions' arrays \cite{Muccino2025}. 
The expected performance of these  libration-free retro-reflectors,  enhanced by the MPAc support, is projected to enable improvements of up to two orders of magnitude in several tests of GR compared to the present centimeter-level accuracy. This augmented capability directly impacts the probes of the weak and strong EP, and guarantees the libration-independent operating conditions necessary to achieve millimeter-level laser-ranging performance.

The analysis of lunar laser ranging data, both historical and next-generation, is performed using the software  $PEP$, developed at the  Center for Astrophysics | Harvard $\&$ Smithsonian since the 1960s \cite{Chandler2021}. Once its modeling and implementation are further refined,  $PEP$ will allow for joint estimation of weak and strong EP violations.  Furthermore, the inclusion of new normal points generated by MPAc allows $PEP$ to extend its scope to tests of extended gravity frameworks (such as MAG,  $f(R)$ models, non-minimal coupled gravity) by incorporating additional dynamical terms directly into the equations of motion and ephemerides, and by performing simultaneous fits against both standard and alternative models~\cite{Bargiacchi2025}.

In addition,  current and forthcoming space missions also provide an ideal laboratory for evaluating  EP-violating effects. The aforementioned MICROSCOPE mission has already established the tightest bound to date on the   E\"{o}tv\"{o}s parameter $\eta $ at the $ 10^{-15}$ level
\cite{MICROSCOPE2022}, a sensitivity that is expected to be improved to $\eta \sim 10^{-18}$ by STEP using superconducting, cryogenic test masses in a drag-free spacecraft \cite{Pereira2016}. Complementing these classical measurements, the European Space Agency STE-QUEST  mission concept proposes to probe the universality of free fall  with an accuracy of $\eta \lesssim 10^{-17}$, relying on quantum matter-wave probes instead of macroscopic test bodies \cite{STE-QUEST2022}. 

In the context of metric-affine gravity, we have seen that the dilation-induced modification of the Newtonian law derived in Eq. \eqref{dilation-MAG-weak-slow} leads to the rescaling factor $\chi \Delta / (4 m_0)$ in the ratio  $m_g/ m_i$, as can be read off from  Eq. \eqref{mg-mi-ratio-MAG-3}.  Future space missions, which span both macroscopic and quantum regimes, together with the  millimeter-level laser-ranging accuracy achievable with MPAc and the enhanced modeling provided by $PEP$, would make it possible to constrain this parameter to extremely small values. This, in turn,   would provide an experimental window on non-metricity signatures,   the associated EP departure, and  their interplay with the thermal effects discussed in Sec. \ref{sec:finite-temperature}.

\section{Einstein Equivalence Principle violation and Fermi-Walker transport in metric-affine geometry} \label{sec:parallel transport and GFW}

In Sec. \ref{Sec:MAG-gauge-theory}, we have explained that  the gauge-theoretic, elementary-matter notion of the Einstein EP can  still be realized pointwise in MAG, while in Sec.  \ref{Sec:pure-dilation},  we have proved how possible departures from  the universality of free fall can emerge  within a pure-dilation regime. In this section, we establish that the modern formulation of the Einstein EP is not valid in general.  As mentioned before, such breakdown can be traced back to the fact that normalization of the metric (as well as the trivialization of  the connection) is a point-dependent gauge choice in MAG (cf. Eqs. \eqref{MAG-orthonormality} and  \eqref{MAG-EP}),  which is  not preserved under parallel transport owing to the non-metricity. From a gauge-theoretic perspective,  this feature is intertwined with departures from local Lorentz invariance in the metric-compatible sense.

To describe this phenomenon,  we introduce  a generalization of the FW derivative to the MAG context. This  novel construction enables a purely geometric and intrinsic analysis of potential violations of the Einstein EP within the most general metric-affine scenario.   The new FW operator shows that  the standard Einstein-EP local Lorentz-frame construction fails  \qm{globally}, i.e.,  along \emph{generic} worldlines, thereby extending the analysis  beyond  the ordinary parallel transport scenario.

We first review the FW derivative in Riemannian geometry  (Sec. \ref{Sec:FW-in-Riemannian}), before turning to its extension to non-Riemannian settings in Secs. \ref{Sec:FW-in-MAG} and \ref{Sec:projection-F}. Finally, we consider possible experimental tests in Sec. \ref{Sec:Exp-test-2}. 

\subsection{Fermi-Walker derivative in Riemannian geometry}\label{Sec:FW-in-Riemannian}

In GR, the FW transport rule defines a prescription for carrying a proper Lorentz frame along an arbitrary timelike worldline $\lambda(s)$  without  inducing a spatial rotation of the spatial axes due to the  acceleration \cite{MTW:gravitation,Hawking:1973uf,Blau:lecturenotes,Bini2002,Poisson:2011motioncurved,romano2019_book_GR,Tsamparlis2025}.  

The local  Minkowski frame attached to an observer  moving on $\lambda(s)$ with  unit  tangent four-velocity  $\dot{\mathbf{x}} \equiv d \mathbf{x}/ds$ is specified via a tetrad   $\{ \tensor{e}{^\mu_a}\}$, which forms a local orthonormal basis  in the tangent space to the spacetime manifold, with $ \tensor{e}{^\mu_{\hat{0}}}$ identified with   $\Dot{x}^\mu$. It is well known that if this frame is  propagated via standard parallel transport along a non-geodesic orbit,   the timelike vector $ \tensor{e}{^\mu_{\hat{0}}}$ will no longer coincide with  $\Dot{x}^\mu$, and  the spatial vectors $ \tensor{e}{^\mu_{\hat{1}}}$, $ \tensor{e}{^\mu_{\hat{2}}}$, and $ \tensor{e}{^\mu_{\hat{3}}}$ will not remain orthogonal to $\Dot{x}^\mu$. On the other hand, if $\{ \tensor{e}{^\mu_a}\}$ is  FW transported along $\lambda(s)$, the resulting frame retains its orthonormality at each point, with $ \tensor{e}{^\mu_{\hat{0}}}=\Dot{x}^\mu$  and the spatial triad undergoing no arbitrary rotation. Physically,  such a frame  can be interpreted as a set of mutually orthogonal, rigid gyroscopes that keep fixed orientation relative to  the observer instantaneous local rest frame while following the path   $\lambda(s)$.

The FW derivative of a vector field $\mathbf{Y}$ along the timelike curve $\lambda(s)$   reads as \cite{Hawking:1973uf}
 \begin{equation} \label{fermi-walker der}
\frac{D^{FW}\tensor{\mathbf{Y}}{}}{ds} = \frac{\Dbol\tensor{\mathbf{Y}}{}}{ds} -\tensor{g}{}\left(\tensor{\mathbf{Y}}{},\tensor{\frac{\Dbol\Dot{\mathbf{x}}}{ds}}{}\right)\tensor{\Dot{\mathbf{x}}}{} +\tensor{g}{}\left(\tensor{\mathbf{Y}}{},\tensor{\Dot{\mathbf{x}}}{}\right)\tensor{\frac{\Dbol\Dot{\mathbf{x}}}{ds}}{} \,,
\end{equation}
where $\Dbol \left(\cdot\right)/ds=   \tensor{\nablabol}{^{}_{\Dot{\mathbf{x}}}} \left(\cdot\right) $ denotes the GR absolute derivative along $\lambda(s)$    and $g\left(\cdot,\cdot\right)$  the scalar product. It is thus evident that the FW derivative boils down to the usual covariant derivative if $\lambda(s)$ is a geodesic. In components, the above formula becomes
\begin{equation} \label{fermi-walker der components}
\dot{x}^\lambda \nablafw_\lambda \tensor{Y}{^\mu} = \dot{x}^\lambda  \tensor{\nablabol}{^{}_{\lambda}} \tensor{Y}{^\mu} + \tensor{F}{^\mu_\nu}\tensor{Y}{^\nu} \,,
\end{equation}
where the skew-symmetric tensor field $\tensor{F}{^\mu_\nu}$ encodes all kinematical information and takes the form
\begin{equation} \label{fermi-walker tensor}
    \tensor{F}{^\mu_\nu} = \tensor{\abol}{^\mu}\tensor{\Dot{x}}{_\nu} - \tensor{\Dot{x}}{^\mu}\tensor{\abol}{_\nu} \,,
\end{equation}
with  $\tensor{\abol}{^\mu}= \dot{x}^\lambda  \tensor{\nablabol}{^{}_{\lambda}} \dot{x}^\mu $ the Riemannian  four-acceleration  orthogonal to $\dot{x}^\mu$.  Thus, the FW covariant derivative operator  (which represents formally  a derivation in the tangent fiber bundles of the vector fields on the spacetime manifold $\mathcal{M}$ \cite{romano2019_book_GR})   is given by
\begin{align}
    \nablafw_\lambda \tensor{Y}{^\mu} = \partial_\lambda Y^\mu +  \tensor{\Gammafw}{^{\mu}_{\lambda\nu}} Y^\nu,
\end{align}
where we have introduced the FW \textit{symbols} 
\begin{equation} \label{F-W symbols}
    \tensor{\Gammafw}{^{\mu}_{\lambda\nu}} = \tensor{\Gammabol}{^{\mu}_{\lambda\nu}} + \left( \tensor{\Dot{x}}{_\nu}\tensor{\nablabol}{_\lambda}\tensor{\Dot{x}}{^\mu} - \tensor{\Dot{x}}{^\mu} \tensor{\nablabol}{_\lambda}\tensor{\Dot{x}}{_\nu} \right) \,.
\end{equation}
These  give rise to a well-defined connection  whose formal definition demands  sophisticated mathematical tools involving  the formalism of second-order tangent bundles (see Ref. \cite{Mian:nullsthesis} for further details).

The FW derivative \eqref{fermi-walker der} satisfies the following key properties:
\begin{itemize}
\item The tangent vector $\Dot{\mathbf{x}}$ to any timelike curve $\lambda(s)$ is FW transported:
\begin{subequations}
\label{FW-derivative-properties}
\begin{align}
\frac{D^{FW} \Dot{\mathbf{x}}}{ds}=0;
\label{FW-property-1}
\end{align}    
\item \text{If $\tensor{\mathbf{X}}{}$ and $\tensor{\mathbf{Y}}{}$ are FW transported along $\lambda(s)$, then}  

\text{their scalar product remains  constant along  $\lambda(s)$}: 
\begin{align}
\frac{d \, g\left(\tensor{\mathbf{X}}{},\tensor{\mathbf{Y}}{}\right)}{ds}=\frac{\Dbol \, g\left(\tensor{\mathbf{X}}{},\tensor{\mathbf{Y}}{}\right)}{ds}=0 ;   
\label{FW-property-2}
\end{align}
\item \text{Vectors $\tensor{\mathbf{Y}}{}$ that satisfy  $g\left(\tensor{\mathbf{Y}}{},\Dot{\mathbf{x}}\right)=0$ along  $\lambda(s)$, re-}

\text{main in the orthogonal subspace to $\Dot{\mathbf{x}}$ and do not}

\text{rotate within it:}
\begin{align}
\frac{D^{FW} \mathbf{Y} }{ds}= \mathop{\mathcal{\stackrel{\circ}{P}}_{\perp }}\nolimits\left(\frac{\Dbol \mathbf{Y} }{ds}\right), 
\label{FW-property-3}
\end{align}
\end{subequations}
$\mathop{\mathcal{\stackrel{\circ}{P}}_{\perp }}\nolimits$ denoting the projection operator in the direction orthogonal to  $\Dot{\mathbf{x}}$, i.e., onto the instantaneous local rest space of the observer (cf. Eq. \eqref{P-perp-GR}, below).
\end{itemize}

The FW derivative can be naturally extended to spacelike curves. Let $\varrho(s)$ be a generic curve with unit tangent vector $\boldsymbol{\tau}$, not necessarily timelike. Then, the FW derivative of a vector field $\tensor{\mathbf{Y}}{}$ along $\varrho(s)$ is
 \begin{equation} \label{fermi walker time or spacelike}
    \frac{D^{FW}\tensor{\mathbf{Y}}{}}{ds} = \frac{\Dbol\tensor{\mathbf{Y}}{}}{ds}
    +\frac{\tensor{g}{}\left(\tensor{\mathbf{Y}}{},\tensor{\frac{\Dbol\boldsymbol{\tau}}{ds}}{}\right)}{\tensor{g}{}\left(\tensor{\boldsymbol{\tau}}{},\boldsymbol{\tau}\right)}\boldsymbol{\tau}
    -\frac{\tensor{g}{}\left(\tensor{\mathbf{Y}}{},\boldsymbol{\tau}\right)}{\tensor{g}{}\left(\boldsymbol{\tau},\tensor{\boldsymbol{\tau}}{}\right)}\tensor{\frac{\Dbol\boldsymbol{\tau}}{ds}}{} \,,
\end{equation}
which reduces to  formula \eqref{fermi-walker der} when $\tensor{g}{}\left(\boldsymbol{\tau},\boldsymbol{\tau}\right)=-1$. 

When the FW derivative is to be evaluated along a congruence of null worldlines with tangent vector $\mathbf{n}$,  a more careful analysis is required. In this context, a complementary null vector $\mathbf{l}$, satisfying $g\left(\mathbf{n},\mathbf{l}\right) \neq 0$, is introduced to span the two-dimensional subspace of vectors normal to  $\mathbf{n}$  \cite{Carroll-book,Poisson-book}. With this auxiliary structure,  it can be shown that the FW derivative of a vector field $\mathbf{Y}$ along the null congruence is
\begin{align} \label{fermi-walker transport for null vec}
    \frac{D^{FW}\mathbf{Y}}{ds}=&\frac{\Dbol\mathbf{Y}}{ds}
    +\frac{g\left(\mathbf{Y},\frac{\Dbol\mathbf{l}}{ds}\right)}{g\left(\mathbf{l},\mathbf{n}\right)}\mathbf{n}
    -\frac{g\left(\mathbf{Y},\mathbf{l}\right)}{g\left(\mathbf{l},\mathbf{n}\right)}\frac{\Dbol \mathbf{n}}{ds} \nonumber\\
    &+ \frac{g\left(\mathbf{Y},\frac{\Dbol\mathbf{n}}{ds}\right)}{g\left(\mathbf{l},\mathbf{n}\right)}{\mathbf{l}}
    - \frac{g\left(\mathbf{Y},\mathbf{n}\right)}{g\left(\mathbf{l},\mathbf{n}\right)}\frac{\Dbol\mathbf{l}}{ds}
    \,.
\end{align}
This construction ensures a well-defined notion of transport for vectors in the transverse subspace. For further details, we refer the reader to Refs.  \cite{Mian:nullsthesis,Geroch:1973am, bargueno:GPHcalc, Bini_2006}.

\subsection{Fermi-Walker derivative in metric-affine geometry}\label{Sec:FW-in-MAG}

Identities  \eqref{FW-derivative-properties} rely crucially on two key features of Riemannian geometry:  the possibility of \textit{always} normalizing the tangent vector $\dot{x}^\mu$ to a timelike path (i.e., $\dot{x}^\mu \dot{x}_\mu=-1$), and  the ensuing orthogonality with the four-acceleration $\abol_\mu$ (i.e., $\dot{x}^\mu \abol_\mu=0$). However, these two conditions generally break down in MAG, as we discussed in Sec. \ref{Sec:MAG-kinematics} (cf. Eqs. \eqref{magnitude 4vel} and \eqref{accel-veloc-product-MAG}). Therefore, in this section we   develop a FW derivative suitable for metric-affine spaces, in order to explicitly account for the non-metricity tensor and its distinctive role. 

Let there be given a curve $\gamma(s)$  with  parameter $s$ and tangent vector $T^\mu$, which satisfies \cite{Agashe:2023vsz}
\begin{equation} \label{magnitude 4vel-bis}
\tensor{T}{^{\mu}_{}}\tensor{T}{^{}_{\mu}} =  \epsilon \phi(\tensor{x}{^\rho}) \,,
\end{equation}
with  $\epsilon = \pm1, 0$  whether $\gamma(s)$ is  spacelike, timelike or null, respectively. For consistency with our notation, in the case of timelike curves, we will denote the tangent vector by $\dot{x}^\mu$, which, we recall, satisfies relation \eqref{magnitude 4vel}.

In his original paper, Fermi considered only vectors orthogonal to $T^\mu$, while Walker  extended this setup  to generic  vectors. A straightforward way to introduce the FW derivative in GR \cite{Mian:nullsthesis,Bini2002}, in line with   the original ideas of Fermi and Walker,   is  to exploit  property \eqref{FW-property-3}, supplemented by the hypotheses that, along $\gamma(s)$: 
\begin{itemize}
\item (i) the length of $T^\mu$ is conserved; 
\item (ii) the scalar product  $g\left(\tensor{\mathbf{Y}}{},\mathbf{T}\right)$ \emph{remains} constant. 
\end{itemize}

These considerations  suggest that a generalized FW derivative can be properly defined  starting from  Eq. \eqref{FW-property-3}, written  in a form  adapted  to MAG settings. Therefore, we introduce the projector  onto the subspace  orthogonal to $\mathbf{T}$:
\begin{align} 
\mathcal{P}_{\perp \;  \beta}^{\; \; \alpha}=\delta^\alpha_\beta -\frac{T^\alpha T_\beta}{\epsilon \phi} \,, \label{projector-MAG}
\end{align}
which allows Eq. \eqref{FW-property-3} to be  re-expressed  as  the  projected-transport condition
\begin{align} \label{GFW-condition-1}
  \mathcal{P}_{\perp} \left[\frac{D}{ds} \left(\mathcal{P}_{\perp} \mathbf{Y}\right)\right] =0 \,,  
\end{align}
where, we recall, $D\left(\cdot\right)/ds$ involves the covariant derivative with respect to the full MAG connection $\tensor{\Gamma}{^{\alpha}_{\mu\nu}}$ (see Sec. \ref{Sec:torsion-non-metricity-curvature}). It is thus clear that  $\mathcal{P}_{\perp}$ is well defined for paths with \(\epsilon\neq 0\), to which we henceforth restrict our attention (a proper definition of the derivative along null worldlines requires a separate procedure,   analogous to that employed in Eq. \eqref{fermi-walker transport for null vec}).  

In view of Eq. \eqref{magnitude 4vel-bis}, it is clear that assumption (i) cannot be retained in MAG. On the other hand, the new FW prescription must enforce (ii) in order for a vector $\mathbf{Y}$ to remain orthogonal to  $\mathbf{T}$ as it is FW transported along $\gamma(s)$. The rationale for this  choice is the following.  FW transport provides, in general, a way to  define, point by point along a  worldline, a tetrad. This scheme necessarily relies on  requirement (ii), which ensures  that the vectors perpendicular to the velocity remain within this orthogonal subspace throughout the evolution. Without  enforcing (ii), any attempt by an observer to carry a tetrad along her/his trajectory would fail from the outset, and the EP would be immediately violated. Nevertheless,  as we will show, even when (ii) is imposed,  the construction of a tetrad along a curve ultimately proves impossible, thereby showing that  in MAG contexts the Einstein EP is inevitably spoiled.

In view of the above considerations, drawing on Eq. \eqref{GFW-condition-1} and exploiting hypothesis (ii) but not (i), we obtain the following generalized FW derivative of the vector $Y^\alpha$ along  $\gamma(s)$:  
\begin{equation} \label{GeneralizedFW components}
    \frac{\mathcal{D}\tensor{Y}{^\alpha}}{ds} =  \frac{D\tensor{Y}{^\alpha}}{ds} -\frac{1}{\epsilon \phi} \tensor{\F}{^\alpha_\beta}\tensor{Y}{^\beta} + \frac{\left(\tensor{T}{_\beta} \tensor{Y}{^\beta} \right)}{\epsilon^2\phi^2}  \left(\tensor{T}{_\lambda}\frac{D \tensor{T}{^\lambda}}{ds}\right)\tensor{T}{^\alpha} \,,
\end{equation}
where  the tensor $    \tensor{\F}{^\alpha_\beta}$ reads as
\begin{align}\label{mathcal-F-tensor}
\tensor{\F}{^\alpha_\beta}= \frac{D\tensor{T}{^\alpha}}{ds}\tensor{T}{_\beta} - \tensor{T}{^\alpha}\frac{D \tensor{T}{_\beta}}{ds} \,,
\end{align}
and, contrary to Einstein gravity, is not antisymmetric  (cf. Eq. \eqref{fermi-walker tensor}; further details on $\tensor{\F}{^\alpha_\beta}$ will be given in Sec. \ref{Sec:projection-F}). We note  that Eq. \eqref{GeneralizedFW components} reduces to Eq. \eqref{fermi walker time or spacelike} in the GR limit,  
where $D\tensor{T}{^\mu}/ds$ and $D\tensor{T}{_\mu}/ds$ become  the contravariant and covariant components of the same vector (whereas in MAG they correspond, for  timelike trajectories, to the proper and anomalous accelerations, respectively; see Eqs. \eqref{proper acceleration in mag} and \eqref{hyper acceleration in mag}),  and the tangent $\tensor{T}{^\mu}$ is orthogonal to its covariant derivative.

The extension of the operator $\mathcal{D}/ds$ to arbitrary-rank tensors can be worked out starting from the following formulas:
\begin{subequations}
\begin{align}
\frac{\mathcal{D}\tensor{Y}{_\mu}}{ds} =&  \frac{D\tensor{Y}{_\mu}}{ds} +\frac{1}{\epsilon \phi} \tensor{\F}{^\alpha_\mu}\tensor{Y}{_\alpha} - \frac{\left(\tensor{T}{_\beta} \tensor{Y}{^\beta} \right)}{\epsilon^2\phi^2}  \left(\tensor{T}{_\lambda}\frac{D \tensor{T}{^\lambda}}{ds}\right)\tensor{T}{_\mu} \,,      
\\
\frac{\mathcal{D}\tensor{A}{^\mu^\nu}}{ds} =&  \frac{D\tensor{A}{^\mu^\nu}}{ds} -\frac{1}{\epsilon \phi}\left(\tensor{\F}{^\mu_\alpha} A^{\alpha \nu}+ \tensor{\F}{^\nu_\alpha} A^{\mu \alpha} \right) 
 \nonumber \\
&+ \frac{1}{\epsilon^2\phi^2} \left(\tensor{T}{_\beta} \frac{D T^\beta}{d s} \right) T_\alpha \left( A^{\alpha \nu} T^\mu + A^{\mu \alpha} T^\nu \right),
\\
\frac{\mathcal{D}\tensor{A}{_\mu_\nu}}{ds} =&  \frac{D\tensor{A}{_\mu_\nu}}{ds} +\frac{1}{\epsilon \phi}\left(\tensor{\F}{^\alpha_\mu} A_{\alpha \nu}+ \tensor{\F}{^\alpha_\nu} A_{\mu \alpha} \right) 
 \nonumber \\
&- \frac{1}{\epsilon^2\phi^2} \left(\tensor{T}{_\beta} \frac{D T^\beta}{d s} \right) T^\alpha \left( A_{\alpha \nu} T_\mu + A_{\mu \alpha} T_\nu \right),
\end{align}
\end{subequations}
from which we derive, in particular,
\begin{align}
 \frac{\mathcal{D}\tensor{g}{_\mu_\nu}}{ds} =&  T^\lambda Q_{\lambda \mu \nu} - \frac{T^\lambda T^\alpha}{\epsilon \phi} \left(T_\mu Q_{\lambda \alpha \nu} + T_\nu Q_{\lambda \alpha \mu}\right)
 \nonumber \\
 &- \frac{2}{\epsilon^2 \phi^2} \left(T_\beta \frac{D T^\beta}{d s}\right) T_\mu T_\nu \,.
 \label{GFW-metric}
\end{align}
It  follows directly from the above relations that $\mathcal{D}/ds$ obeys the  Leibniz rule and boils down to the  ordinary derivative when acting on scalar functions.

In the case of  timelike curves, for which  $T^\mu \equiv \dot{x}^\mu$, with $\dot{x}^\mu \dot{x}_\mu = -\phi$ (cf. Eq. \eqref{magnitude 4vel}),  the FW derivative of the four-velocity field yields
\begin{subequations}
\label{GFW-velocity-1-and-2}
\begin{align}
\frac{\mathcal{D}\tensor{\dot{x}}{^\alpha}}{ds}  & = \frac{\dot{\phi}}{\phi} \dot{x}^\alpha \,,
\label{GFW-velocity-1}
\\
\frac{\mathcal{D}\tensor{\dot{x}}{_\alpha}}{ds}  & = 0 \,,
\label{GFW-velocity-2}
\end{align}
\end{subequations}
where we have exploited Eqs. \eqref{proper acceleration in mag}--\eqref{accel-veloc-product-MAG}. In other words, the tangent vector to $\gamma(s)$ is not FW-conserved, unlike in Riemannian geometry.

It is worth noting that an alternative  construction of  the  FW derivative  can be provided,  which can be viewed as the \qm{dual} of the operator in Eq. \eqref{GeneralizedFW components}. This second version  can be derived by means of an equivalent approach, where one first evaluates the generalized FW derivative on vectors belonging to the subspace  orthogonal to $\mathbf{T}$, and then extends its action to generic vectors. In this way, we obtain
\begin{align} \label{GeneralizedFW components-B2}
\frac{\tilde{\mathcal{D}}\tensor{Y}{^\mu}}{ds} =&  \frac{D\tensor{Y}{^\mu}}{ds} -\frac{1}{\epsilon \phi} \tensor{\F}{^\mu_\alpha}\tensor{Y}{^\alpha}  - \frac{\left(\tensor{T}{_\beta} \tensor{Y}{^\beta} \right)}{\epsilon^2\phi^2}  \left(\tensor{T}{^\lambda}\frac{D \tensor{T}{_\lambda}}{ds}\right)\tensor{T}{^\mu} \,,
\end{align}
and we find that the FW derivative $\tilde{\mathcal{D}}/ds$  is tantamount  to $\mathcal{D}/ds$, in the sense that it yields the \qm{dual} relations of  those in Eq. \eqref{GFW-velocity-1-and-2}:
\begin{subequations}
\begin{align}
 \frac{\tilde{\mathcal{D}}\tensor{\dot{x}}{^\alpha}}{ds}  &=0,
 \\
 \frac{\tilde{\mathcal{D}}\tensor{\dot{x}}{_\alpha}}{ds}  &=   \frac{\dot{\phi}}{\phi} \dot{x}_\alpha.
\end{align}
\end{subequations}

The two definitions \eqref{GeneralizedFW components} and \eqref{GeneralizedFW components-B2} coincide on the subspace orthogonal to $T^\mu$ and  differ only by the last  term proportional to $1/(\epsilon^2 \phi^2)$; obviously, both reduce to the standard FW transport in the metric-compatible regime. The possibility of setting up these two  alternatives  is fully consistent  with the outcome  of Ref.  \cite{Manoff_1998}, which points out that in generic metric-affine spacetimes different, yet  equally valid, constructions of a generalized FW derivative can be formulated. However, our definitions \eqref{GeneralizedFW components} and \eqref{GeneralizedFW components-B2} have the key strength of being  built from the same antisymmetric tensor \eqref{mathcal-F-tensor}, unlike what typically occurs  in  MAG \cite{Manoff_1998}.  

Having clarified these points, we hereafter employ the derivative \eqref{GeneralizedFW components} and examine its basic properties. By construction, the analogue of Eq. \eqref{FW-property-3} is satisfied, while we have already proved that the counterpart of Eq. \eqref{FW-property-1} does not hold in MAG (cf. Eq. \eqref{GFW-velocity-1-and-2}). Likewise,  if $\mathcal{D}Y^\mu/ds=\mathcal{D}X^\mu/ds=0$, then we find that
\begin{align}
\frac{d}{ds}&\left(g_{\mu \nu} Y^\mu X^\nu\right)
\nn \\
&=Q_{\lambda \mu \nu} T^\lambda Y^\mu X^\nu -\frac{ Q_{\lambda \alpha \mu}}{\epsilon \phi} T^\lambda T^\mu\Biggl[\left(T_\beta Y^\beta\right) X^\alpha
\nn \\
&+ \left(T_\beta X^\beta\right)Y^\alpha\Biggr]
-\frac{2}{\epsilon^2 \phi^2}    \left(X_\mu T^\mu\right) \left( Y_\nu T^\nu\right) T_\lambda \frac{D T^\lambda}{ds},
\label{property-GFW-scalar}
\end{align}
which is not surprising, as the corresponding  GR property \eqref{FW-property-2} relies on the antisymmetry of $\tensor{F}{^\mu_\nu}$,  a feature that is lost in MAG (see Eq. \eqref{mathcal-F-tensor}). A further crucial  aspect,   reflecting the basic requirements of the new FW construction, is that if $\mathcal{D}Y^\mu/ds=0$, then
\begin{align}
 \frac{d}{ds}\left(g_{\mu \nu} Y^\mu T^\nu\right)  =0, 
\end{align}
which means that if the vector $Y^\mu$  is  initially perpendicular  to the tangent $T^\mu$,  FW transport along $\gamma(s)$ preserves this orthogonality during  the entire evolution. However, a caveat should  be mentioned at this point: in deriving the above relations  we assumed that the contravariant components of a generic vector have vanishing FW derivative. The conclusions change if one instead imposes the condition on the covariant components (e.g., $\mathcal{D}Y_\mu/ds=0$),  due to fact that the metric tensor is not   FW-conserved   (see Eq. \eqref{GFW-metric}).

As a consequence of the above formulas,   the generalized FW scheme  \eqref{GeneralizedFW components}  fails to provide a consistent prescription for defining a tetrad \footnote{Here the observer-adapted tetrad should not be confused with the  frame field, dual of the coframe  $\boldsymbol{\vartheta}^a= \tensor{e}{_\mu^a} \boldsymbol{\dd} x^\mu$ introduced in Sec. \ref{Sec:MAG-gauge-theory}. Although they are objects of the same geometrical type, the former is a choice of local frame along a specific worldline, whereas the latter is  not tied to a specific observer, and is part of the metric-affine gauge structure.} $\{\mathbf{e}_{\hat{0}}=\dot{\mathbf{x}},\mathbf{e}_{\hat{1}},\mathbf{e}_{\hat{2}},\mathbf{e}_{\hat{3}}\}$  that remains orthonormal everywhere on the worldline $\gamma(s)$ of an observer. This obstruction is caused by two fundamental issues: the FW derivative of $\mathbf{e}_{\hat{0}}$ is not zero (see Eq. \eqref{GFW-velocity-1}), and  the orthogonality between two spatial vectors $\mathbf{e}_{\hat{i}}$ is not preserved throughout $\gamma(s)$, owing to Eq. \eqref{property-GFW-scalar}. 

The impossibility of propagating an orthonormal tetrad via the FW procedure represents a clear geometric indication of the intrinsic breakdown of the Einstein EP in generic metric-affine backgrounds.  Indeed, it is known that in Einstein theory a local manifestation of the EP is provided by Riemann and Fermi normal coordinates   \cite{Manasse1963,MTW:gravitation,Hartle2003-book} (see also Ref. \cite{Battista2020} for a recent application), where the latter can be defined along an arbitrary  worldline by resorting to the ordinary FW transport of the tetrad   \cite{MTW:gravitation,Nesterov1999,Poisson:2011motioncurved}.  This  standard construction, however,  fails  in MAG arena since, as we have shown above,  the FW scheme prevents  the establishment of an orthonormal frame that  renders the metric locally Minkowskian on the observer trajectory. Our analysis thus demonstrates that  the EP violation is not confined to the pure-dilation dynamics   examined  in Sec. \ref{Sec:pure-dilation}, but is a distinctive feature of the broad MAG framework. 

Some final remarks are in order. First, it is straightforward to show that the EP breaking can be equivalently described via  the \qm{dual} operator $\tilde{\mathcal{D}}/ds$. Second, we notice that generalized notions of FW transport in non-Riemannian settings have appeared in the literature, see e.g.  Refs. \cite{Manoff_1998,Kocharyan:2004dz,Gurzadyan2017,Tsamparlis2025} for further details. However, our projector-based  construction differs in both method and emphasis: it does not assume metric compatibility, is tailored to MAG,  and naturally   addresses the question of EP noncompliance.

\subsection{Projection structure of the tensor $ \tensor{\mathcal{F}}{^\alpha_\beta}$ }\label{Sec:projection-F}

In this section, we further analyze the properties of the tensor $ \tensor{\mathcal{F}}{^\alpha_\beta}$  characterizing the generalized FW derivative (see Eqs. \eqref{GeneralizedFW components} and \eqref{mathcal-F-tensor}). We first consider the Riemannian scenario  in Sec. \ref{subsubsec: analysis in V4}, and then turn to  MAG  in Sec. \ref{subsubsec:analysis in L4}.

\subsubsection{Riemannian spacetime} \label{subsubsec: analysis in V4}

In a Riemannian spacetime, let there be given a  timelike curve, and   introduce the projection operators
\begin{subequations}
\label{projectors-Riemann}
\begin{align}
{\tensor{\PparllGR}{^\alpha_\mu}} =& - \dot{x}^\alpha \dot{x}_\mu  \,, 
 \\
{\tensor{\PperpGR}{^\alpha_\mu}} =& \delta^\alpha_\mu +\dot{x}^\alpha \dot{x}_\mu  \,,\label{P-perp-GR}
\end{align}
\end{subequations}
which allow    any tensorial quantity to be decomposed into components parallel and orthogonal to the tangent vector $ \dot{x}^\mu$ to the curve. Thus, the projection of the Riemannian acceleration along the tangent direction satisfies
\begin{equation} \label{parallel proj riemann}
{\tensor{\PparllGR}{^\alpha_\mu} \frac{\Dbol\dot{x}^\mu}{ds} =0 } \,,
\end{equation}
which, in turn, yields
\begin{equation}\label{perp proj riemann}
    {\tensor{\PperpGR}{^\alpha_\mu}} \frac{\Dbol\dot{x}^\mu}{ds} =  \frac{\Dbol\dot{x}^\alpha}{ds}\;.
\end{equation}

In view of Eq. \eqref{parallel proj riemann}, we  now examine the  tensor $ \tensor{F}{^\alpha_\beta}$  entering  the Riemannian FW derivative  \eqref{fermi-walker der components}. Bearing in mind Eq. \eqref{fermi-walker tensor}  and performing a double parallel projection, we get 
\begin{align} \label{projection FW tensor}
    &\tensor{\PparllGR}{^\mu_\alpha}\tensor{\PparllGR}{^\beta_\nu}\tensor{F}{^\alpha_\beta} =  \tensor{\PparllGR}{^\mu_\alpha} \left[(-\dot{x}^\beta\dot{x}_\nu) \abol^\alpha \dot{x}_\beta  \right.
    \nonumber \\
      & \left. - (-\dot{x}^\beta\dot{x}_\nu) \dot{x}^\alpha\abol_\beta  \right] = \tensor{\PparllGR}{^\mu_\alpha} \left(\abol^\alpha \dot{x}_\nu \right) = 0 \,,
\end{align}
showing that   $ \tensor{F}{^\alpha_\beta}$ possesses no component along the longitudinal direction  and thus lies entirely in the subspace orthogonal to $\dot{x}^\mu$. 

The above   results  rely crucially on the metric-compatibility condition $ \tensor{\nablabol}{^{}_{\lambda}}g_{\mu \nu}=0$ and the ordinary normalization $\dot{x}^\mu \dot{x}_\mu =-1$. With these arguments  in mind, we  now extend the discussion to the wider context of  non-Riemannian geometry.

\subsubsection{Metric-affine spacetime } \label{subsubsec:analysis in L4}

As we set out before, the local violation of the Lorentz invariance in metric-affine spacetime is related to the presence  of the non-metricity tensor \eqref{non-metricity tensor} (or, equivalently, the one-form $\boldsymbol{Q}_{ab}$, see Sec. \ref{Sec:gauge-affine-group}), which entails varying-magnitude vectors and gives rise to  two different accelerations, namely the proper  and the anomalous one (cf. Eqs. \eqref{proper acceleration in mag}--\eqref{hyper acceleration 2}). Introducing  the orthogonal and longitudinal projectors along the lines of Eq.   \eqref{projectors-Riemann} (cf. Eq. \eqref{projector-MAG}), and considering a timelike curve  with tangent vector $\dot{x}^\mu$ satisfying the relation \eqref{magnitude 4vel}, the tangent projection of the proper  acceleration yields
\begin{equation} \label{parallel proj proper-acc}
    {\tensor{\Pparll}{^\alpha_\mu} \frac{D\dot{x}^\mu}{ds} = \frac{1}{2} \dot{x}^\alpha} f_+ \,,
\end{equation}
while  for the anomalous contribution we obtain
\begin{equation} \label{parallel proj hyper-acc}
    {\tensor{\Pparll}{^\mu_\alpha} \frac{D\dot{x}_\mu}{ds} = \frac{1}{2}\dot{x}_\alpha }f_-\,,
\end{equation}
where
\begin{equation}
    f_{\pm}{\tiny(x^\rho)} := \frac{\dot{\phi}{\tiny(x^\rho)}}{\phi{\tiny(x^\rho)}} \pm \frac{\dot{x}^\lambda\tensor{Q}{_{\lambda\mu\nu}}{\tiny(x^\rho)}\dot{x}^\mu\dot{x}^\nu}{\phi{\tiny(x^\rho)}} \,,
\end{equation}
the orthogonal projections following immediately from $\tensor{\Pperp}{^\alpha_\mu} =  \delta^\alpha_\mu - \tensor{\Pparll}{^\alpha_\mu} $. 

Thus, unlike  the Riemannian case, where the acceleration $\abol^\mu$ satisfies Eq. \eqref{parallel proj riemann},  in MAG both $A^\mu$ and $\tilde{a}_\mu$ carry in general a nonvanishing component  along the tangent direction, as expected from Eq. \eqref{accel-veloc-product-MAG}.  For this reason, while in the Riemannian framework  one can always uniquely decompose a generic vector along the mutually orthogonal directions defined by $\abol^\mu$ and $\dot{x}^\mu$, in MAG such a construction is no longer available, as neither $A^\mu$ nor $\tilde{a}_\mu$ are  perpendicular to the four-velocity. This  different kinematical setup inexorably leads to the new FW derivative introduced in Eq. \eqref{GeneralizedFW components}, where the Riemannian tensor $\tensor{F}{^\mu_\nu}$ is  replaced by $\tensor{\F}{^\mu_\nu}$. 

If we decompose  the accelerations according to 
\begin{align}
A^\alpha &= \left(\tensor{\Pperp}{^\alpha_\mu} + \tensor{\Pparll}{^\alpha_\mu}\right)A^\mu, 
\nonumber \\
\tilde{a}_\alpha &= \left(\tensor{\Pperp}{_\alpha^\mu} + \tensor{\Pparll}{_\alpha^\mu}\right)\tilde{a}_\mu,
\end{align}
and  use Eq. \eqref{mathcal-F-tensor}, then  Eqs. \eqref{parallel proj proper-acc} and \eqref{parallel proj hyper-acc} imply that $\tensor{\mathcal{F}}{^\alpha_\beta}$ admits the decomposition
\begin{align} \label{generalised FW tensor}
\tensor{\mathcal{F}}{^\alpha_\beta} =& A^\alpha\dot{x}_\beta - \dot{x}^\alpha \tilde{a}_\beta = (\tensor{\Pperp}{^\alpha_\mu}A^\mu + \tensor{\Pparll}{^\alpha_\mu}A^\mu)\dot{x}_\beta 
\nonumber\\  
&- \dot{x}^\alpha (\tensor{\Pperp}{_\beta^\mu}\tilde{a}_\mu + \tensor{\Pparll}{_\beta^\mu}\tilde{a}_\mu) =:\tensor{\Fperp}{^\alpha_\beta} + \tensor{\Fparll}{^\alpha_\beta}  \,,
\end{align}
where
\begin{subequations}
\label{F-perp-and-parall}
\begin{align}
\tensor{\Fperp}{^\alpha_\beta}&:= \tensor{\Pperp}{^\alpha_\mu}A^\mu\dot{x}_\beta - \tensor{\Pperp}{_\beta^\mu}   \tilde{a}_\mu \dot{x}^\alpha, 
\\
\tensor{\Fparll}{^\alpha_\beta}&:= \frac{1}{2}(f_+ - f_-)\dot{x}^\alpha \dot{x}_\beta=\frac{\dot{x}^\rho\tensor{Q}{_{\rho\sigma\tau}}\dot{x}^\sigma\dot{x}^\tau}{\phi}\dot{x}^\alpha \dot{x}_\beta.
    \end{align}
\end{subequations}
We thus acknowledge a difference with the Riemannian relation \eqref{projection FW tensor}, as $ \tensor{\mathcal{F}}{^\alpha_\beta}$ now contains also a longitudinal component apart from the orthogonal one. In the Riemannian scenario, the former vanishes identically, while  $ \tensor{\Fperp}{^\alpha_\beta}$ boils down to $ \tensor{F}{^\alpha_\beta}$ owing to Eq. \eqref{perp proj riemann}.

Interestingly, in the special case where the manifold, equipped with a particular non-metricity, admits the so-called Yano-Schr\"{o}dinger connection  \cite{Schrodinger:spacetime,Klemm2020,Csillag:2024eor}, then the geometry satisfies
\begin{subequations}
\begin{align}
\tensor{Q}{_{(\lambda\mu\nu)}} &= 0, \\
\dot{\phi}&=0,\\
\dot{x}_\mu \abol^\mu &= 0,
\end{align}
\end{subequations}
which imply that only $\tensor{\Fperp}{^\alpha_\beta}$ contributes to $\tensor{\F}{^\alpha_\beta}$, as the longitudinal part $\tensor{\Fparll}{^\alpha_\beta}$ is identically zero (see Eqs. \eqref{generalised FW tensor} and \eqref{F-perp-and-parall}).

\subsection{Possible experimental tests}\label{Sec:Exp-test-2}

In the previous sections, we have seen how in generic metric-affine backgrounds the standard GR notion of propagating a locally orthonormal tetrad along an observer worldline fails, and hence the usual construction of a local Minkowski frame (as well as the local special-relativistic interpretation of nongravitational physics) is obstructed. This setup provides an intrinsic and geometric indication that the Einstein EP is spoiled.

When the non-metricity tensor is treated, on laboratory scales, as a weak and slowly varying \emph{background} field, its  presence effectively introduces preferred directions in the spacetime for freely falling observers, leading to orientation- and boost-dependent signatures in local measurements \cite{Foster2016}. Despite differing from the approach presented in Refs. \cite{Obukhov2023-proc,Obukhov2024a}, this mechanism offers a direct link between non-metricity and possible departures from  local Lorentz invariance. A context where this scenario is systematically addressed is furnished by the standard-model extension, an effective field theory originally motivated by potential Lorentz-breaking phenomenological effects in string theory \cite{Colladay1996,Kostelecky2003}. This framework is based on a Lagrangian containing all parameterized coordinate-independent Lorentz-violating terms that can be formed from known fields \cite{Colladay1996,Kostelecky2003}. 

Recently, it has been argued that a background non-metricity can induce an effective Lorentz violation in a laboratory, even when the underlying gravitational theory with non-metricity is locally Lorentz invariant \cite{Foster2016}. Building on this claim,  a correspondence can be established between the spacetime non-metricity and the operators responsible for Lorentz non-invariance in the standard-model extension. In this way, existing experimental bounds on effective Lorentz-symmetry breaking can be translated into laboratory constraints on non-metricity components with high sensitivity \cite{Foster2016}. These bounds restrict viable metric-affine models and may be interpreted as indirect constraints on Einstein-EP departures induced by non-metricity.

Among the most relevant probes of local Lorentz invariance worth mentioning are modern Michelson-Morley-type experiments, first devised in Ref. \cite{Muller2003}, where the isotropy of the speed of light is explored by comparing the frequencies of orthogonal cryogenic optical resonators over extended periods of time, exploiting the Earth rotation.
Such measurements are naturally interpreted within the standard-model-extension pattern and can be mapped onto bounds on Lorentz-violating coefficients, which in turn can be translated into restrictions on the non-metricity. 

More generally, precision tests of Lorentz symmetry can therefore serve as high-precision probes of non-metricity. These include  clock-comparison (Hughes-Drever-type \cite{Kostelecky2018}) experiments  \cite{Qin2025}, space-clock missions such as ACES \cite{Cacciapuoti2024}, resonant-cavity experiments \cite{Muller2003b,Zhang2021}, spin-precession measurements \cite{Smiciklas2011}, atom-interferometric tests \cite{Chung2009,Gao2024}, and, more recently, space-based gravitational-wave detectors \cite{Qin2023}. Together, these experimental platforms provide a complementary and powerful avenue for assessing the validity of Einstein EP within MAG.

\section{Discussion and conclusions} \label{Sec:Conclusion}

In this paper,  we have investigated possible EP violations in MAG. Specifically, we have shown that universality of free fall can fail, and we have provided a proof of the breakdown of the Einstein EP in its modern formulation.  

First, we have outlined the main geometrical and gauge-theoretic features of MAG  in Sec. \ref{sec:MAG}. Then,  in Sec.  \ref{sec:finite-temperature}, we have explained how thermal radiative corrections in finite-temperature QFT and classical effective gravity models spoil the EP by entailing a mismatch between  the gravitational and inertial masses of a point particle evolving in a Riemannian background. We have then  proved that these effects admit a  natural geometric analogue in the broader MAG framework.
 
Within the wide range of  scenarios permitted by MAG,  Sec. \ref{Sec:pure-dilation} demonstrates that the dynamics of a test body possessing a dilation charge offers a particularly clear setting for probing  deviations from  geodesic worldlines. The dilation current introduces an additional term in the equations of motion,  resulting in a  rescaling of the Newtonian law in Eq. \eqref{dilation-MAG-weak-slow},  which remarkably  exhibits a close similarity to the Riemannian  result \eqref{radial motion at finite t}. The ensuing expressions \eqref{mg-mi-ratio-MAG-1}--\eqref{mg-mi-ratio-MAG-3} attain the same functional form as the finite-temperature formulas \eqref{inertial mass}--\eqref{EP at finite T} for the ratio $m_g/m_i$,  although their origin is different. As pointed out in  Sec. \ref{Sec:experiments},  the  combined use of laser-ranging techniques and the software $PEP$ could enable  a test of these EP departures.

Section \ref{sec:parallel transport and GFW}  highlights,  \emph{in full generality}, a direct geometric signature of the Einstein EP breakdown intrinsic to MAG by defining a novel FW derivative that   extends  the  Riemannian FW  transport prescription. 

The Einstein EP is usually stated in two ways. Formulation (i) corresponds to the gauge-theoretic, elementary-matter, and purely pointwise realization of the principle relevant to minimally coupled  matter, and  expresses the local reduction of nongravitational physics to special relativity in suitable (anholonomic) frames. In version (ii), instead, the Einstein EP is understood in modern terms as the conjunction of the weak EP, local Lorentz invariance, and local position invariance. While these notions are tantamount in standard metric theories of gravity, they need not be so in more general frameworks such as metric-affine gravity, where only  (i) is retained because one can trivialize the metric and affine connection at a single event.

From a gauge-theoretic perspective, this situation stems from the extension of the gauge group  from the Lorentz group $SO(1,3)$ to the general linear group $GL(4,\mathbb{R})$, as part of the affine gauge group $A(4,\mathbb{R})$. This scenario  is associated with  departures from local Lorentz invariance and  the emergence of non-metricity. As a consequence, orthonormal frames can still be selected at any given spacetime event, but they cannot, in general, be extended along worldlines while preserving orthonormality,  which  becomes a local-at-a-point gauge choice  in MAG (cf. Eq. \eqref{MAG-orthonormality}). In contrast,  in metric-compatible theories  the Einstein EP is realized smoothly in  local Lorentz frames at each  point, as orthonormal  tetrads  are preserved under parallel or FW transport. In view of the analysis presented this paper, we can conclude that the local Lorentz-frame realization underlying (ii)  is spoiled  not only under parallel transport, but also via the new FW transport.

The  FW operator introduced here  is indeed  tailored to MAG and  permits  proving that an observer cannot propagate an orthonormal tetrad basis $\{\mathbf{e}_{a}\}$ along his/her trajectory for two main reasons:  $\mathbf{e}_{\hat{0}}$ does not coincide with the four-velocity owing to Eq.  \eqref{GFW-velocity-1};   the spatial vectors $\mathbf{e}_{\hat{1}}, \mathbf{e}_{\hat{2}},$ and $\mathbf{e}_{\hat{3}}$ fail to remain mutually orthogonal, as expressed by Eq. \eqref{property-GFW-scalar}.  

To the best of our knowledge, the new  FW scheme  \eqref{GeneralizedFW components} has not appeared in this form in the literature and thus can open new avenues for research on gravity in non-Riemannian geometries. The potential applications of this derivative operator, as well as the EP departures emerging in the motion of a generic hypermomentum-charged test particle in non-Riemannian backgrounds, are topics that deserve to be further examined.

%--------------------------------------
\section*{Acknowledgements}
The authors acknowledge the support of  Istituto Nazionale di Fisica Nucleare (INFN), as part of the MoonLIGHT-2 experiment in the framework of the research activities of the {\it Commissione Scientifica Nazionale No. 2 (CSN2)}.   Authors  are grateful to  Marcello Miranda for fruitful discussions about non-metricity; to Riccardo March for  the  critical reading of the  preliminary version of the manuscript and for the useful comments provided;   to Anish Agashe for the discussion on various types of curves within the metric-affine context.  SC thanks also  the Gruppo Nazionale di Fisica Matematica (GNFM)  of Istituto Nazionale di Alta Matematica (INDAM) for the support. The authors thank the anonymous referee for the useful referee report.  
%--------------------------------------

%--------------------------------

\bibliography{references}

\end{document}